\documentclass[%
 reprint,
 amsmath,amssymb,
 aps,
]{revtex4-2}

\usepackage{graphicx}
\usepackage{dcolumn}
\usepackage{bm}
\usepackage{float}
\usepackage{placeins}
\usepackage{xcolor}
\newcommand{\sgn}{\operatorname{sgn}}

\begin{document}

\preprint{APS/123-QED}

\title{Behavior of Ising spins and ecological oscillators on dynamically rewired small-world networks.}

\author{Davi Arrais Nobre}
\email{dv.nobr@gmail.com}
\affiliation{Department of Physics and Astronomy, University of California, Davis, California, 95616, USA}

\author{Karen C. Abbott}
\affiliation{
Department of Biology, Case Western Reserve University, Cleveland, Ohio, 44106, USA
}%

\author{Jonathan Machta}
\affiliation{Department of Physics, University of Massachusetts, Amherst, Massachusetts, 01003, USA}
\affiliation{Santa Fe Institute, 1399 Hyde Park Road, Santa Fe, New Mexico 87501, USA}%

\author{Alan Hastings}%
\affiliation{Department of Environmental Science and Policy, University of California, Davis, California, 95616, USA}
\affiliation{Santa Fe Institute, 1399 Hyde Park Road, Santa Fe, New Mexico 87501, USA}

\date{\today}

\begin{abstract}
Many ecological populations are known to display a cyclic behavior with period 2. Previous work has shown that when a metapopulation (group of coupled populations) with such dynamics is allowed to interact via nearest neighbor dispersal in two dimensions, it undergoes a phase transition from disordered (spatially asynchronous) to ordered (spatially synchronous) that falls under the 2-D Ising universality class. While nearest neighbor dispersal may satisfactorily describe how most individuals migrate between habitats, we should expect a small fraction of individuals to venture on a journey to further locations. We model this behavior by considering ecological oscillators on dynamically rewired small-world networks, in which at each time step a fraction $p$ of the nearest neighbor interactions is replaced by a new interaction with a random node on the network. We measure how this connectivity change affects the critical point for synchronizing ecological oscillators. Our results indicate that increasing the amount of long-range interaction (increasing $p$) favors the ordered regime, but the presence of memory in ecological oscillators leads to quantitative differences in how much long-range dispersal is needed to order the network, relative to an analogous network of Ising spins. We also show that, even for very small values of $p$, the phase transition falls into the mean-field universality class, and argue that ecosystems where dispersal can occasionally happen across the system's length scale will display a phase transition in the mean-field universality class.
\end{abstract}

\maketitle


\section{\label{sec:intro}Introduction}
Spatial synchrony of dynamical systems is a well-known phenomenon that has been studied in many disciplines \cite{strogatz-sync, liebhold2004, kuramoto-osc}. In ecology, synchrony can be achieved between ecological oscillators, which are often modeled using discrete time equations that display a stable period-2 cycle. In such case, spatial synchrony means that populations at different locations reach high population in the same generation, and low population in the following, such that they are positively correlated. Different processes may cause such spatial order. For instance, long-range environmental correlations may drive the coupled populations, which in ecology are referred to as metapopulations, to a similar state, through a phenomenon known as the Moran effect \cite{moran1953, goldwyn2011}, but short range interactions such as dispersal among neighboring habitat patches can also be sufficient to create long-range order \cite{goldwyn2011, noble2015}. Understanding the mechanism behind spatial synchrony is of crucial importance in ecology since it may affect the stability of some species \cite{abbott2011}. 

Environmental stochasticity may disrupt spatial synchrony even in the presence of some dispersal between habitat patches \cite{goldwyn2008, noble2015}. By increasing the dispersal rate relative to the noise level, locally coupled ecological oscillators will undergo a continuous phase transition from disorder (spatially asynchronous) to order (spatially synchronous) that has been shown to be in the 2-D Ising universality class \cite{noble2015}. Experimental data for a 2-dimensional pistachio orchard suggest that nearest-neighbor interactions are enough to generate long-range order, and support the computational result that the phase transition falls into the 2-D Ising universality class \cite{noble2018}. Experimental data also show, in this case for a 1-dimensional ring lattice, that nearest neighbor dispersal is not sufficient for maintaining an ordered state, but introducing occasional random long-range dispersal significantly increases the correlation between populations that are further apart along the ring \cite{fox2018}. A similar result has been derived for the XY model in one dimension, in which the introduction of a few long-range interactions led to the emergence of an ordered state at finite temperature \cite{kim2001}.

In this work, we compare the Ising model and ecological oscillators with the introduction of occasional, random long-range coupling. This corresponds to a generalization of the model from Noble et al.\ \cite{noble2015}, and was inspired by work done on small-world networks \cite{strogatz1998, kim2001, herrero2002, hong2002}. We simulate ecological populations on networks that range from a regular 2-dimensional square lattice to a random graph with average degree equal to that of the square lattice ($\langle z\rangle=4$). Ultimately, we are interested in the intermediate case, in which most of the interactions are with nearest neighbors, and a small fraction of sites interact with random nodes on the network that are not nearest neighbors. We compare the results for the ecological oscillators with those for Ising spins with occasional long-range interactions.

Section \ref{sec:network} details the network structure used for both the ecological oscillators and Ising spins, while section \ref{sec:dynamics} presents the rules that govern each system, section \ref{sec:results} presents our results, which are discussed in section \ref{sec:discussion}, and the concluding remarks are in section \ref{sec:conclusion}.

\section{\label{sec:network}Dynamically rewired 2-D small-world networks}
Each ecological oscillator or Ising spin is represented by a node, and nodes interact with others if they are connected by an edge. Most of the work done so far on ecological oscillators assumes that such connections are either to nearest neighbors only \cite{noble2015, noble2018, araujo2008, durrett1994-1}, exist evenly over a wide neighborhood \cite{durrett1994-1, durrett1999}, exist at random between any two nodes \cite{johst2002}, or exist between all pairs of nodes \cite{marti2003, roos1998}. In simple terms, the first case assumes the connectivity between patches can be described by a 2-dimensional regular lattice, while the second can be described by a square lattice with connections between the $k$ nearest neighbors $(k>>1)$, the third by a random graph, and the fourth by a fully connected network. 

In real life ecosystems with annual reproduction and dispersal, it is reasonable to assume that the majority of a species' migration happens to neighboring habitat patches, but some individuals occasionally venture on a longer journey to distant patches. Examples span a wide variety of systems, such as trees whose seeds may be carried by the wind, aquatic larvae that may be carried by water currents, or insects that inhabit trees and may fly to distant trees. To describe this behavior, we replace a fraction $p$ of the edges in a 2-dimensional square lattice (with only nearest neighbor connections) by random connections, in a process similar to the one proposed by Watts and Strogatz \cite{strogatz1998} and that has been used to study both Ising spins \cite{herrero2002} and other types of oscillators \cite{hong2002}. Note, however, that such models use a stationary network, while ours {\color{black
}starts with a regular underlying structure that at every iteration is rewired, introducing long-range connections that only exist at each iteration of the system's dynamics.}

We start our network as a regular undirected 2-dimensional square lattice with side length $L$, such that the total number of nodes is $N=L^2$, with edges connecting nearest neighbors only. We study systems of size $L=16$, $32$, $64$, and $128$,  with periodic boundaries. To introduce occasional long-range interaction, we randomly select $N'$ edges to be rewired, where $N'$ is a random number drawn from a binomial distribution with $2N$ trials and probability of success (rewiring) $p$, which yields an expected value $\langle N'\rangle = 2Np$. If an edge connecting nodes $j$ and $k$, with $j<k$, is selected to be rewired, it is replaced by a new edge connecting nodes $j$ and $k'$, where $k'$ is chosen at random from any other node on the network, provided that we are not recreating the previous edge nor introducing a double edge. 

\begin{figure}[ht]
    \centering
    \includegraphics[width=1\linewidth]{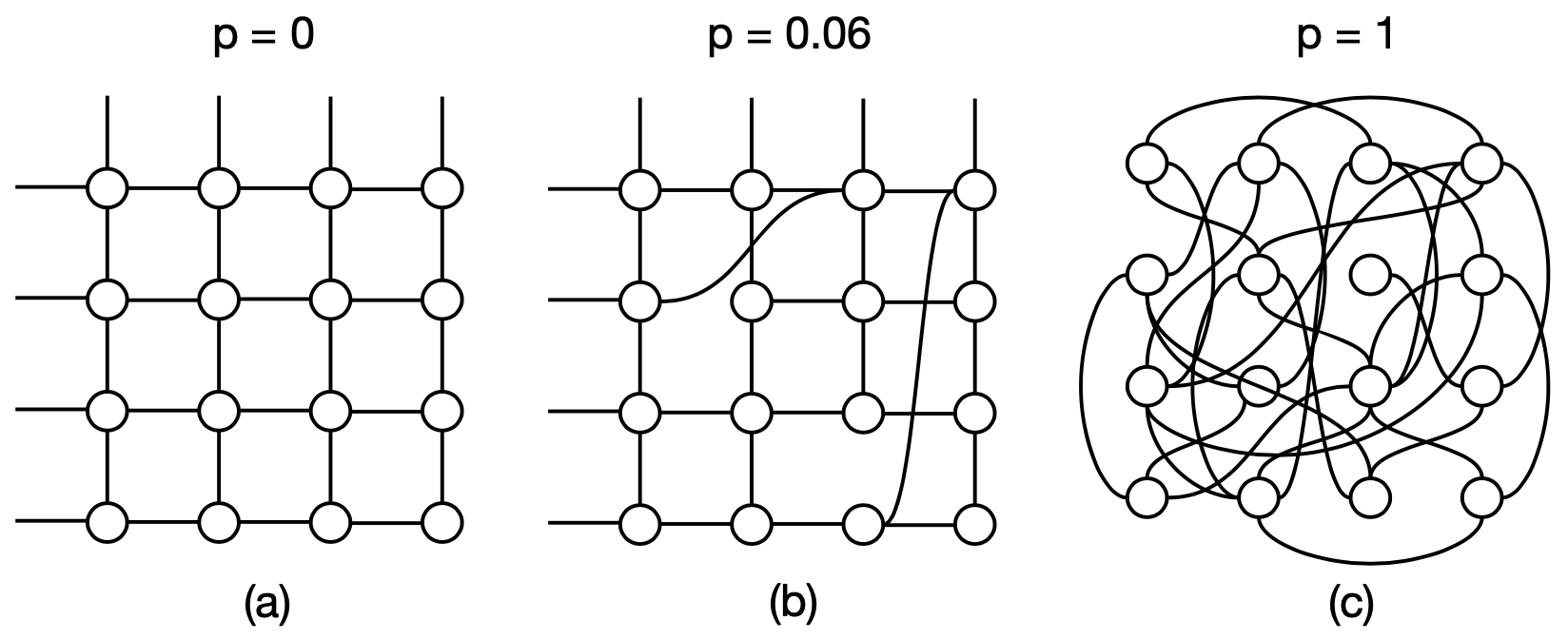}
    \caption{\label{fig:networks} From left to right: a regular 2-D square lattice (a) with periodic boundary conditions, a small-world network (b) created from the regular square lattice with $p=0.06$, and a random graph (c) with average degree $\langle z\rangle = 4$, which corresponds to rewiring every edge present in the square lattice ($p=1$).}
\end{figure}

With this process, the total number of edges on the network is conserved, but a fraction $p$ of them now connects nodes that are not nearest neighbors and are potentially very far apart in physical space, while a fraction $1-p$ of the edges connects nearest neighbors. Note that the average degree remains $\langle z\rangle=4$, but now the degree distribution has non-zero variance. Keeping the number of edges constant is important in order to measure solely the effect of the long-range interaction, since effectively increasing the average degree of the network would naturally tend to synchronize the system. Figure \ref{fig:networks} depicts the two limiting cases, $p=0$ and $p=1$, and one intermediate case, $p=0.06$.

The rewiring process is inspired by ecological dispersal and is  distinct from the physically motivated quenched and annealed cases of small-world networks. Rewiring {\color{black
}always} starts from a regular 2-dimensional square lattice, such that each iteration of the system's dynamics is performed with a different network generated with the same rewiring probability $p$. In more colloquial terms, $p$ is a measure of the amount of long-range interactions, but these connections do not always happen between the same two pairs of nodes. In ecological terms, this means that individuals migrating to a distant location choose that location at random, which is a reasonable assumption given that such long-range dispersal often happens due to changing air or water currents, or simply due to diffusive motion of individuals. Similarly, nodes that are physically nearest neighbors will be connected most of the time (for small $p$), but not all the time. Rewiring the network at the beginning of every iteration,  is a key difference between our model and most literature examples, which usually use quenched networks and need to perform several simulations to average over the ensemble of possible different networks \cite{herrero2002, hong2002, kim2001}. Our rewiring process is also distinct from an annealed small-world network, since rewiring is done disregarding the current state of the nodes to be rewired (see section \ref{subsec:ising}).

\section{\label{sec:dynamics}Dynamics}
\subsection{\label{subsec:oscillator}Ecological oscillators}
To simulate each ecological oscillator, we use a first order difference equation, in which the population density at node $j$ at time $t+1$ depends on the population density at the same node at time $t$, that is to say $X_{j,t+1}=f(X_{j,t})$. Many non-linear models of population dynamics display a cyclic regime with period two (see \cite{abbott-enc} for some examples). In this work, we use the Ricker map with the addition of noise.  The noisy Ricker dynamics of oscillator $j$, before accounting for dispersal, is given by, 
\begin{equation}
    \label{eq:ricker}
    f(X_{j,t}) = X_{j,t}e^{r(1-X_{j,t})}e^{\sigma\xi_{j,t}}.
\end{equation}
The first exponential term in Eq.\ \eqref{eq:ricker} is the non-linear growth rate of the deterministic Ricker map \cite{ricker1954, ricker-enc}, which displays a two-cycle behavior when $2<r<2.5$ (see \cite{strogatz-nonlinear} for a proof method). The second exponential is the noise term \cite{abbott-enc}, where $\xi_{j,t}\sim \mathcal{N}(0,1)$ and $\sigma$ is the noise level, which is the control parameter we choose for the system.  Adding noise accounts for environmental differences through time and between habitat patches and better reproduces the noisy oscillations that are observed in real life cases. The Ricker map has been chosen out of many possible options due to its simplicity and convenience, since having exponential functions for growth and noise eliminates the possibility of the nonphysical case of a negative population. It has been shown that, while the functional form of the population dynamics may affect some quantities related to the phase transition, it does not affect the qualitative aspects of the phenomenon nor does it alter the universality class \cite{noble2015}.

Equation \eqref{eq:ricker} describes the local population dynamics at each node $j$, we still need to consider how different nodes interact via dispersal. If node $j$ is connected to $z_j$ other nodes, we assume a fraction $\epsilon$ of the population leaves $j$ through each edge. Similarly, the same fraction $\epsilon$ enters node $j$ coming from nodes connected to $j$ so that,
\begin{equation}
    \label{eq:model}
    X_{j,t+1} = (1-z_j\epsilon )f(X_{j,t})+\epsilon \sum_{k\in \mathcal{N}_j}f(X_{k,t}),
\end{equation}
where $f(X_{j,t})$ refers to the noisy Ricker map given in Eq.\ \eqref{eq:ricker}, the $(1-z_j\epsilon)$ term accounts for the population leaving node $j$, and the summation is over nodes $k$  connected to $j$ and accounts for the population moving into node $j$. We use $\epsilon = 0.025$ throughout this paper, such that an average of $10\%$ of each node's population migrates at each time step. This value allows for an easier comparison with previous results \cite{noble2015}. Note that $\epsilon$ or some combination of $\epsilon$ and noise level,  $\sigma$ could be chosen as the control parameter of the system. Any of these choices are expected to yield the same qualitative behavior and universality class for the transition to synchrony.  

The dynamics of the system is described by Eq.\ \eqref{eq:model}.  until stationary behavior is reached, and then performing additional iterations to collect data. We always perform local deterministic growth first, followed by multiplication by the noise term, and finally by dispersal, as depicted in equations \eqref{eq:ricker}-\eqref{eq:model}, but the order in which these processes are done does not qualitatively affect the results \cite{vahini2020}.

\begin{figure}[ht]
    \centering
    \includegraphics[width=0.9\linewidth]{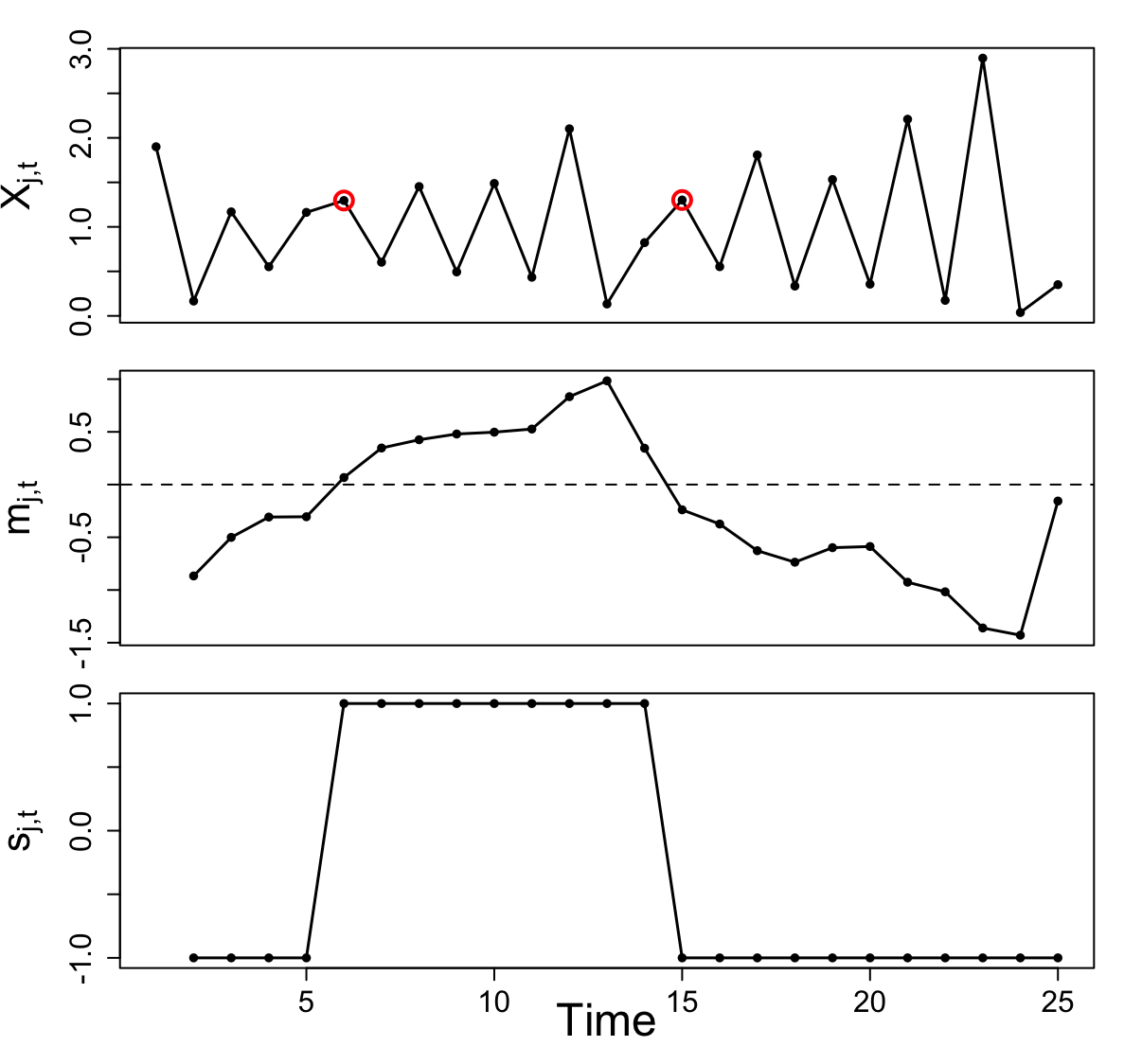}
    \caption{\label{fig:phase} Respectively from top to bottom: local population (Eq.\ \eqref{eq:ricker}), two-cycle variable (Eq.\ \eqref{eq:phase-variable}), and phase variable (analogous to the Ising spin value). Red circles on the top plot mark the times at which the phase of oscillation changes, due to the population either increasing or decreasing at consecutive time steps. Note that ecological oscillators differ from Ising spins in that $m_{j,t}$ is an unbounded continuous variable. One could reduce the dynamics of the ecological oscillators to the corresponding Ising spins (as is done in the bottom panel), but significant information is lost in that transformation \cite{vahini2020}.}
\end{figure}

At each time step of the simulation, the square lattice is rewired and all nodes are updated following Eq.\ \eqref{eq:model}. We repeat this process for $10^7$ time steps, and the last $8 \cdot 10^6$ times steps are used for stationary state measurements. The ordering of the system is determined by the phase of oscillation of each individual node. Since the local dynamics is a noisy period-2 cycle, the population at each node is expected to be high at even time steps, or at odd time steps. We define a two-cycle variable for each node at each time step, $m_{j,t}$, as the alternating-sign first difference in population
\begin{equation}
    \label{eq:phase-variable}
    m_{j,t} = (-1)^t\frac{X_{j,t}-X_{j,t-1}}{2},
\end{equation}
such that $m_{j,t}$ is positive if the population is high at even time steps and negative if the population is high at odd time steps. This two-cycle variable contains both amplitude and phase information, with the amplitude being given by the absolute value of $m_{j,t}$, and the phase being given by its sign. The variable $s_{j,t}=\sgn (m_{j,t})$ is called the phase variable. It may change due to the presence of noise, and will be the analogous of an Ising spin (although the order parameter is calculated using $m_{j,t}$, as shown in Eq.\ \eqref{eq:order-osc}). Figure \ref{fig:phase} shows the relationship between $X_{j,t}$, $m_{j,t}$, and $s_{j,t}$, which is the phase of oscillation and will correspond to the Ising spin.

The order parameter at time $t$, as defined in \cite{noble2015}, is 
\begin{equation}
    \label{eq:order-osc}
    m^E_t = \dfrac{1}{N}\left|\sum_{j=1}^N m_{j,t}\right|.
\end{equation}
Note that $m^E_t\approx 0$ if each node on the network oscillates independently, and $m^E_t>0$ if most nodes are in phase. To determine the stationary order parameter we average $m^E_t$ over a long period of time. The absolute value in Eq.\ \eqref{eq:order-osc} is added to minimize the effects of the system-wide phase shifts that happen close to criticality. 

To determine the critical noise that separates the ordered and disordered phase, we use the Binder cumulant method \cite{binder1981, landaubinder}, where the Binder cumulant, $U$ is defined as,
\begin{equation}
    \label{eq:binder}
    U = 1-\dfrac{\langle m^4\rangle}{3\langle m^2\rangle^2},
\end{equation}
and the expected values are taken from the distribution of $m_t$ in the stationary state. 
{\color{black
}The crossing point of Binder cumulant curves for different system sizes are known to be a highly reliable estimator of the critical point \citep{binder1981}.}

\subsection{\label{subsec:ising}Ising Model}
Here, the dynamics of the Ising spins are governed by the standard Metropolis algorithm \cite{Newmanbook}. Each node $k$ has a spin $s_k$ that takes two possible values, $+1$ or $-1$, and the energy functional, or Hamiltonian, of the system is given by 
\begin{equation}
    \label{eq:hamiltonian}
    H = -\dfrac{J}{2}{\color{black
    }\sum_{j=1}^{N}}\sum_{k\in \mathcal{N}_j} s_k s_j,
\end{equation}
where $J$ is assumed to be $1$ without loss of generality and the sum is over nodes $k$ connected to node $j$. The factor of $2$ is to eliminate double counting. The control parameter for the Ising spins is the temperature $T$, which plays the same role as $\sigma$ for ecological oscillators. At  temperature $T$, a spin can flip with probability
\begin{equation}
    \label{eq:spinflip}
    P_{\text{flip}} =
    \left\{ 
    \begin{aligned}
        & 1, \text{ if } \Delta H<0 \\
        & e^{-\frac{\Delta H}{k_B T}}, \text{ if } \Delta H>0
    \end{aligned}
    \right.
    ,
\end{equation}
where $\Delta H$ is the change in energy associated with the spin flip, and we take Boltzmann's constant to be one, $k_B = 1$.  At each time step of the simulation, the square lattice is rewired following the process described in Sec.\ \ref{sec:network} and then $N$ spins are chosen, with replacement, and sequentially flipped according to Eq.\ \eqref{eq:spinflip}. 

\begin{figure}[ht]
    \centering
    \includegraphics[width=0.9\linewidth]{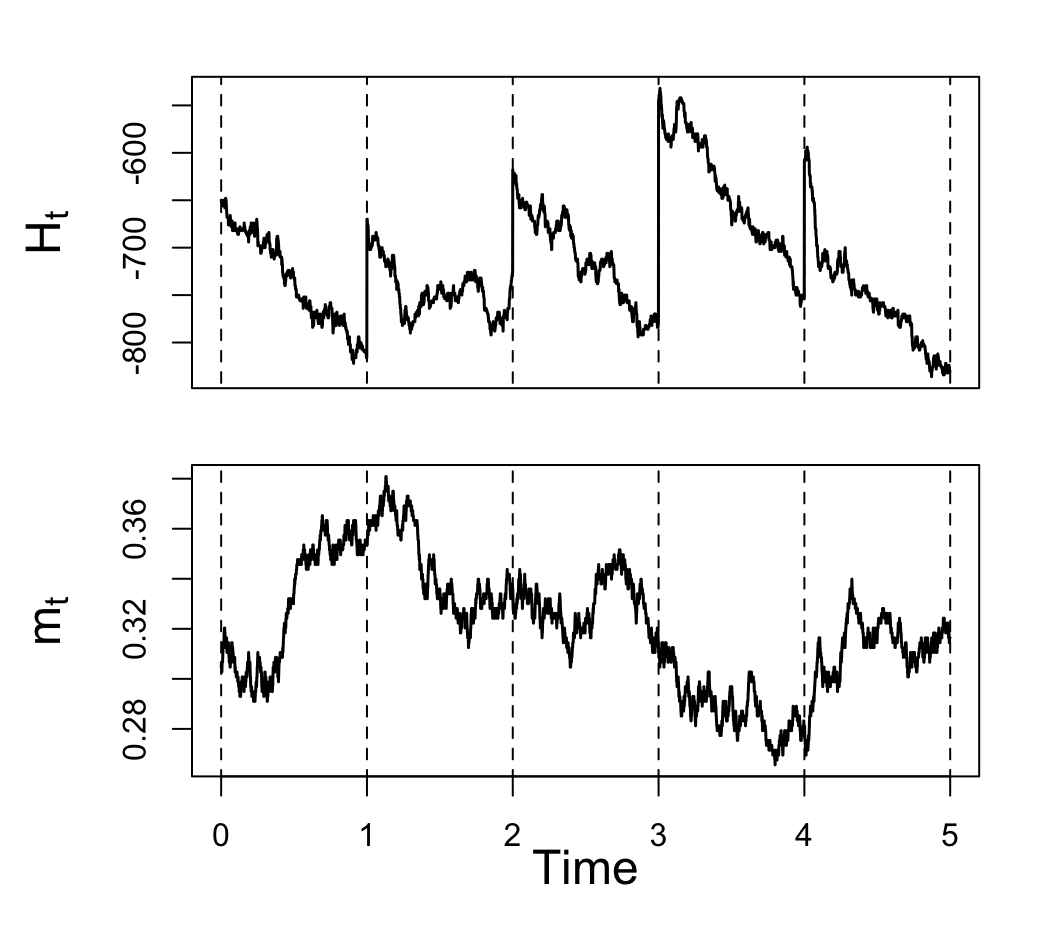}
    \caption{\label{fig:balance} 
    Energy (top, equation \eqref{eq:hamiltonian}) and order parameter (bottom, equation \eqref{eq:order-ising}) behavior in the stationary state for a network with $N=1024$ Ising spins ($L=32$) and rewiring probability $p=0.2$. The dashed lines indicate the beginning of each time step, when rewiring is performed, and the behavior in between is from the sequential spin flips ruled by equation \eqref{eq:spinflip}. }
\end{figure}

For $p>0$, since the rewired edges are accepted with no regards to the spin state, the energy of the system will typically increase after rewiring, and relax towards its equilibrium value after spin flips are performed. This violates detailed balance and the system approaches a stationary state but does not achieve thermal equilibrium. Nonetheless, we find that the magnetic properties of the system after a long time are equilibrium-like. Figure \ref{fig:balance} depicts this behavior for a network with $L=32$ and $p=0.2$ in the stationary state, showing that the energy jumps discontinuously higher after rewiring and then relaxes toward its equilibrium value for that rewiring. Note that the average height of the energy spikes divided by the temperature corresponds to the entropy production of the system, and can therefore be understood as a measure of how far the system is from equilibrium due to rewiring. On the other hand, the order parameter is continuous at rewiring and its time series appears to be time reversal invariant.

This process is repeated for a total of $5\cdot 10^6$ time steps, and data from the final $4\cdot 10^6$ time steps is used for stationary state measurements. The order parameter at each time step is given by 
\begin{equation}
    \label{eq:order-ising}
    m^I_t = \frac{1}{N}\left|\sum_{j=1}^N s_{j,t}\right|,
\end{equation}
and the critical temperature for each value of rewiring probability $p$ is determined using the  Binder cumulant method.

\section{\label{sec:results}Results}
\begin{figure*}[ht]
\includegraphics[scale=0.2175]{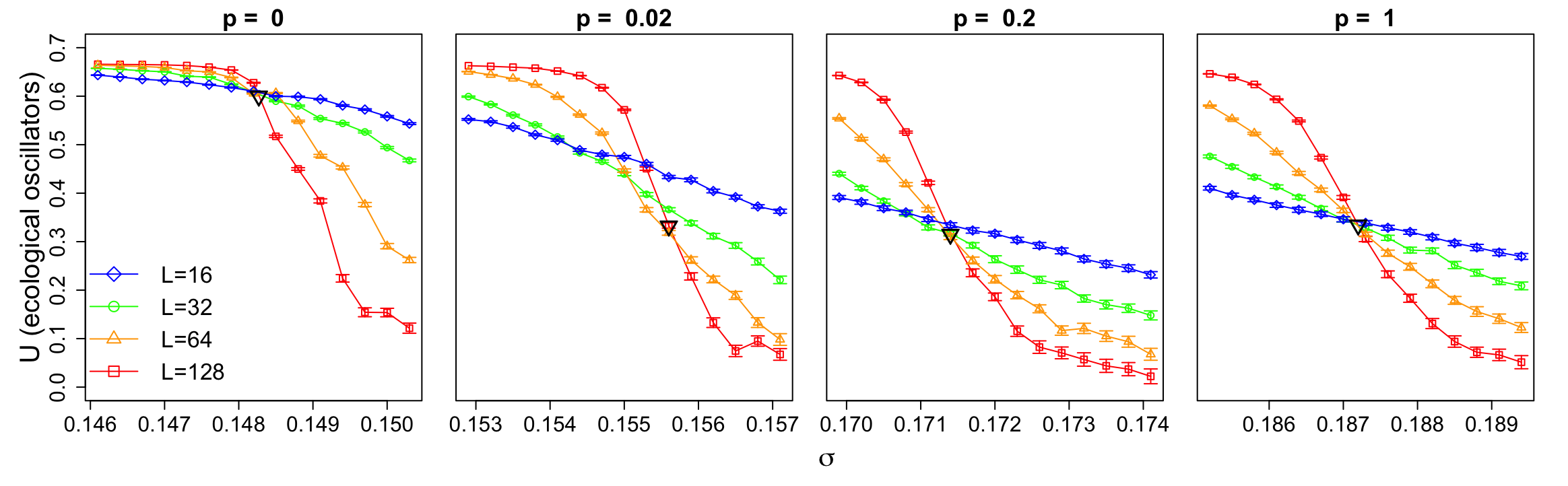}
\includegraphics[scale=0.2175]{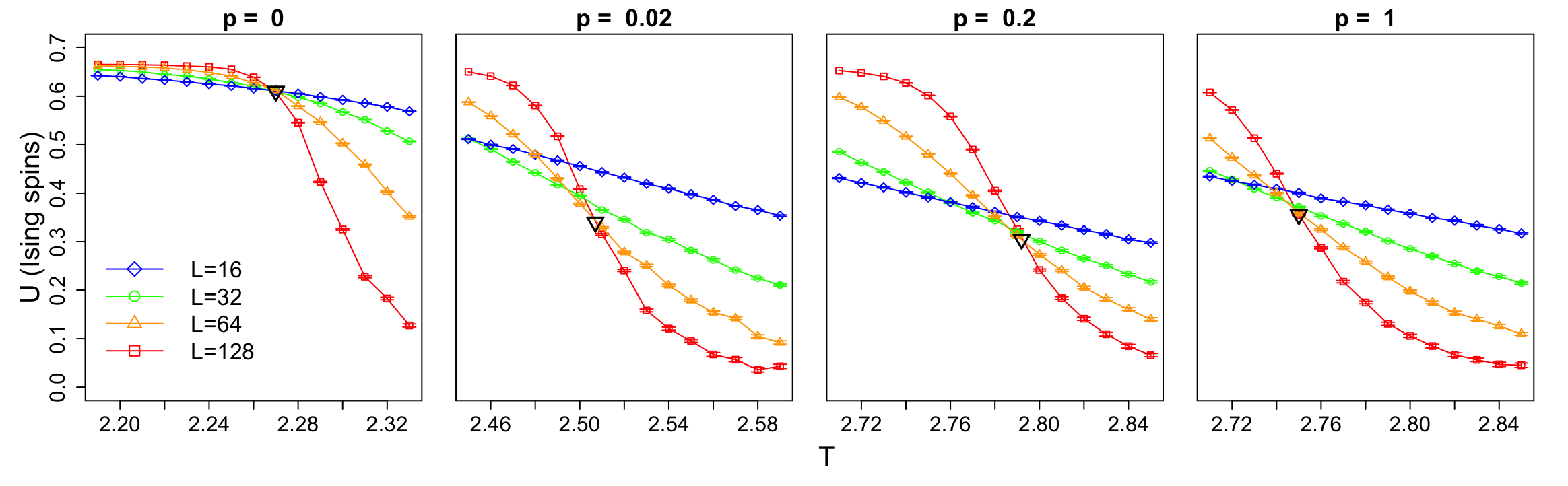}
\caption{\label{fig:binder-osc} Binder cumulant plots for ecological oscillators (top row) and Ising spins (bottom row) for different values of rewiring probability $p$. Error bars are small and were calculated using the bootstrap method. {\color{black
}This method consists of resampling the data with replacement, averaging them, and taking the standard deviation of the averages after a suitably large number of resamples \cite{Newmanbook}}. The solid black triangle indicates the crossing point used to determine the critical control parameter. When the curves did not all cross at the same point, the crossing of the curves corresponding to the two largest systems was used (see the $p=0.02$ plot). The Binder cumulant values in the ordered and disordered phases are the same for the ecological oscillators as for Ising spins: $U=2/3$ in the ordered phase, and $U=0$ in the disordered phase \cite{binder1981}.}
\end{figure*}

For both the ecological oscillators and the Ising spins, we analyzed the stationary state of the system on networks of four different sizes, $L=16$, $32$, $64$, and $128$, and thirteen values of rewiring probability, eleven evenly spaced points between $p=0$ and $0.2$ and two points for the higher values $p=0.6$ and $1$. The critical point for each value of $p$ was determined using the Binder cumulant, which has a value approximately independent of the system size at criticality \cite{binder1981, landaubinder}, although finite size effects exist and are discussed in the next section. Binder cumulant curves for different rewiring probabilities $p$ are shown in Fig.\ \ref{fig:binder-osc} as a function of the control parameter.

\begin{figure*}[!ht]
\includegraphics[scale=0.55]{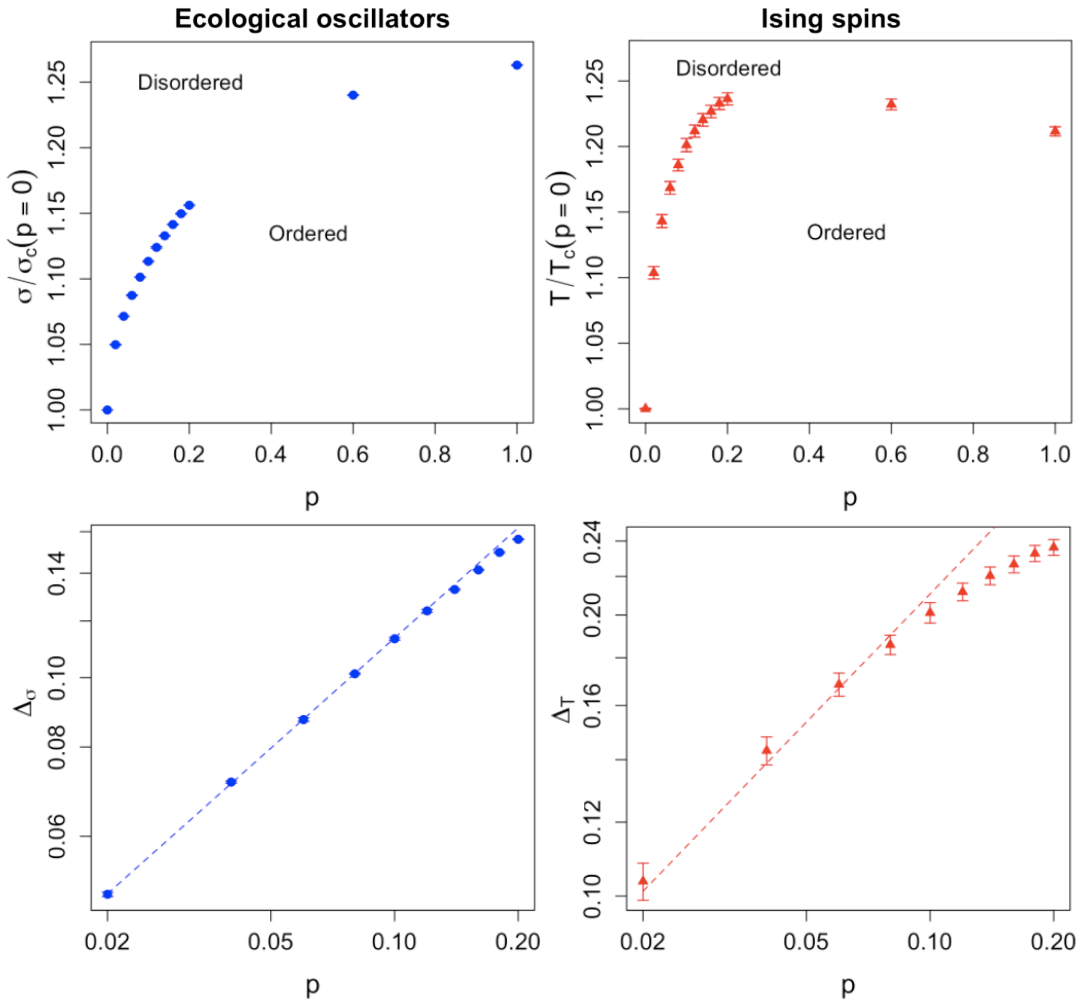}
\caption{\label{fig:phase-diagram} {\color{black
}The points in the top plots indicate the critical control parameter for each value of $p$. For ecological oscillators (left), the critical noise increases monotonically as we increase $p$, with decreasing slope. For Ising spins (right), the critical temperature increases more rapidly, reaching a maximum value at $p=0.2$ and then decreasing slightly for higher values of $p$. The reason for this decrease is discussed in the next section. Error bars were propagated from the uncertainties in the Binder cumulant measures (Fig.\ \ref{fig:binder-osc}) following the process in \cite{taylor-book}. The error bars for ecological oscillators are smaller than the data points. The bottom plots show the change in the critical control parameter (Eq.\ \eqref{eq:critical-change}), which increases with $p$ according to a power law for low values of $p$. For ecological oscillators, the exponent of such power law is $0.510\pm 0.005$, while for Ising spins it is $0.46\pm 0.02$. The dashed lines show the best power law fit using the four points with $p\in [0.02,0.08]$ for ecological oscillators, and the three points with $p \in [0.02,0.06]$ for Ising spins. The power law fit and uncertainties were calculated following the process in \cite{num-recipe}.}}
\end{figure*}

The phase diagrams showing the separation between the ordered and disordered phases are in Fig.\ \ref{fig:phase-diagram}. On these graphs, the vertical axis displays the control parameter divided by its critical value for a regular lattice ($p=0$), so a direct comparison between the two different systems is straightforward. {\color{black
}The critical control parameters obtained for the regular lattice were $T_c(p=0)=2.265\pm 0.002$ for Ising spins, consistent with the exact solution, and $\sigma_c(p=0) = 0.1483 \pm 0.0009$ for ecological oscillators},  consistent with the literature \cite{ noble2015, vahini2020}. The bottom panels in Fig.\ \ref{fig:phase-diagram} show, on log-log plots, the change in the critical control parameter for ecological oscillators and Ising spins with $p$ relative to the $p=0$ case. The vertical axis in these plots is given by 
\begin{equation}
\label{eq:critical-change}
    \Delta_f = \dfrac{f_c(p)-f_c(p=0)}{f_c(p=0)},
\end{equation}
where $f=\sigma$ or $T$ for ecological oscillators and Ising spins, respectively. 

\begin{figure}[ht]
\includegraphics[scale=0.18]{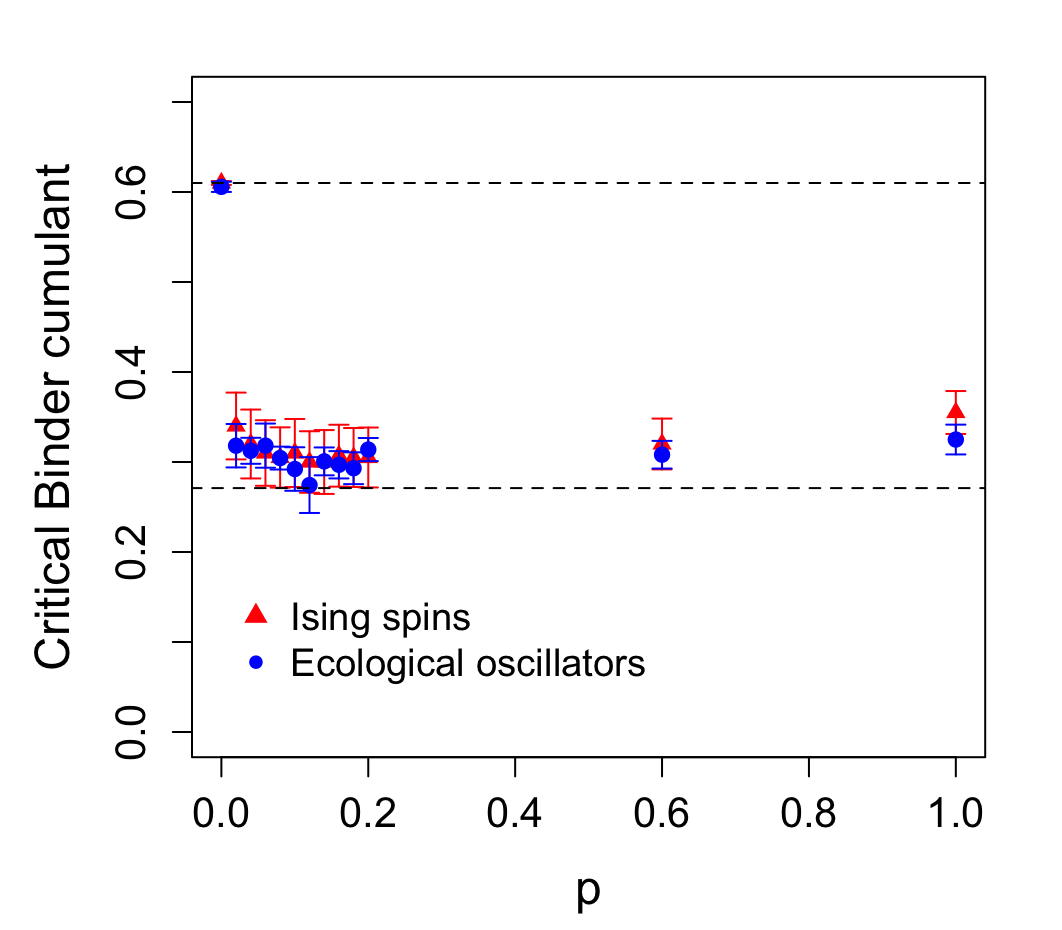}
\caption{\label{fig:critical-binder} The critical Binder cumulant displays the same behavior for both Ising spins and ecological oscillators. For $p=0$, both have a value consistent with the 2-D Ising universality class \cite{janke1994, kastening2013} (top dashed line), but for $p$ as low as $0.02$, both systems show a critical Binder cumulant much closer to the one consistent with the mean-field value, $U\approx 0.27$  \cite{parisi1996} (bottom dashed line).}
\end{figure}

Current literature indicates that the critical value of the Binder cumulant is constant within a universality class provided that the system is isotropic \cite{chen2004}, and it is equal to approximately $0.61$ in the 2-D Ising universality class \cite{janke1994, kastening2013}, and $0.27$ in the mean-field universality class \cite{parisi1996}. Figure \ref{fig:critical-binder} shows a plot of these values for each rewiring probability for both systems. Even for a rewiring probability as low as  $p=0.02$ the critical Binder cumulant indicates that the systems are apparently in the mean-field universality class.

\begin{figure}[ht]
\includegraphics[scale=0.5]{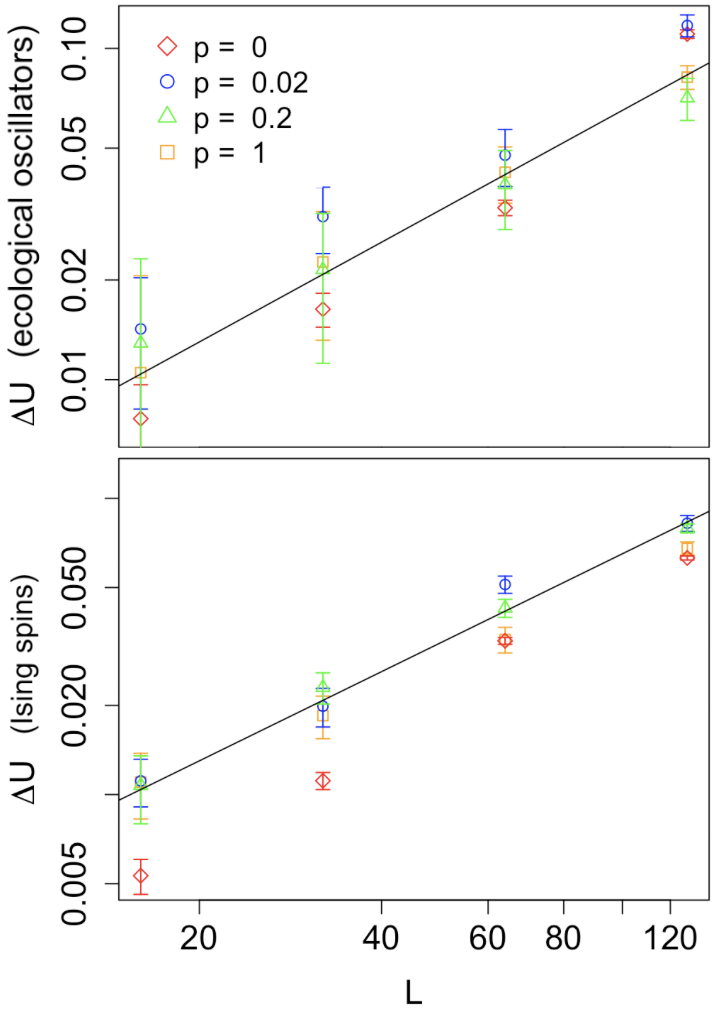}
\caption{\label{fig:binder-finite} Finite-size scaling of the Binder cumulant shows a power law increase on its derivative with the linear size of the system. Ecological oscillators are shown on the top plot and Ising spins at the bottom. The line in both graphs shows a power law with exponent $1$ for reference.}
\end{figure}

While the Binder cumulant is approximately size-independent, finite-size scaling indicates that the derivative of $U$ with the control parameter close to criticality scales with the system size according to 
\begin{equation}
\label{eq:binder-scaling}
    \Delta U = U(T_1)-U(T_2) \propto L^{1/\nu}, 
\end{equation}
for $T_1$ and $T_2>T_1$ both sufficiently close to $T_c$ \cite{landaubinder, kim2001}, where $\nu$ is the critical exponent of the correlation length. Figure \ref{fig:binder-finite} shows this power-law scaling. For both systems and all values of $p$, we found $\nu=1$ within error bars. The apparent discrepancy between this value and the mean-field exponent ($\nu =1/2$) is addressed in the next section.

Now that we know how the critical point shifts with $p$, we can proceed to analyze the effects of rewiring on the order parameter.  Due to the increase in the critical control parameter caused by rewiring, shown in Fig.\ \ref{fig:phase-diagram}, it is useful to study the order parameter in terms of the reduced control parameter, which measures how far away we are from the critical point for each value of rewiring probability. Using $f$ to represent either $\sigma$ or $T$, the reduced control parameter, $f_r$, is given by
\begin{equation}
\label{eq:reduced-temperature}
    f_r = \dfrac{f-f_c(p)}{f_c(p)},
\end{equation}
where $f_c(p)$ is the critical value for a given $p$. Using this metric, we see that, while increasing rewiring tends to increase the critical control parameter, it decreases the order parameter close to criticality.

\begin{figure*}[ht]
\includegraphics[scale=0.2175]{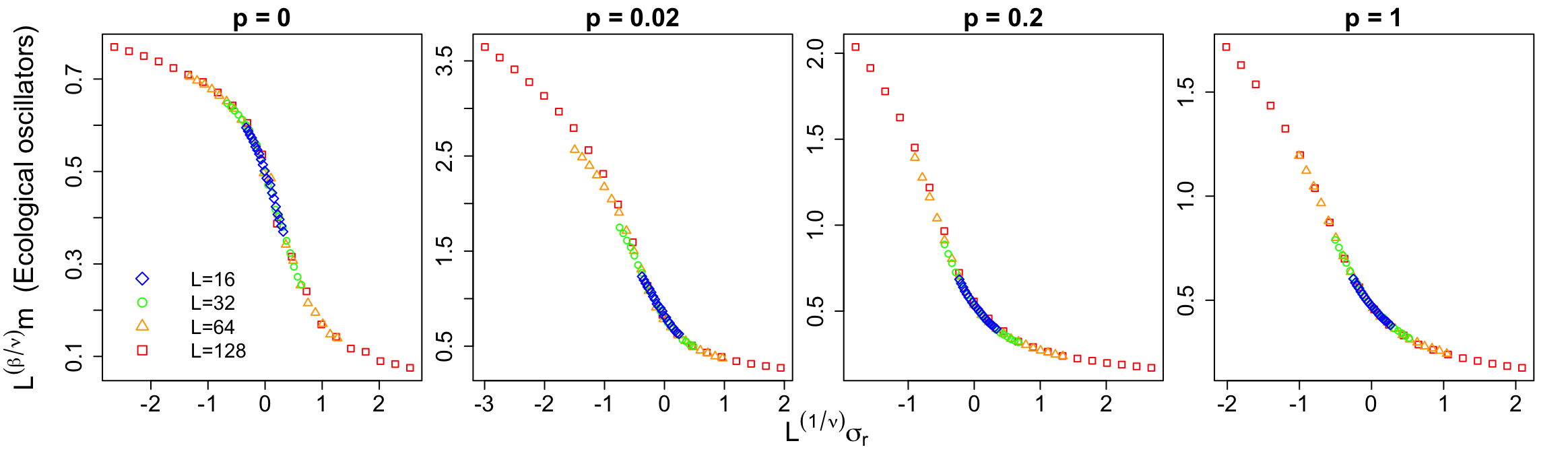}
\includegraphics[scale=0.2175]{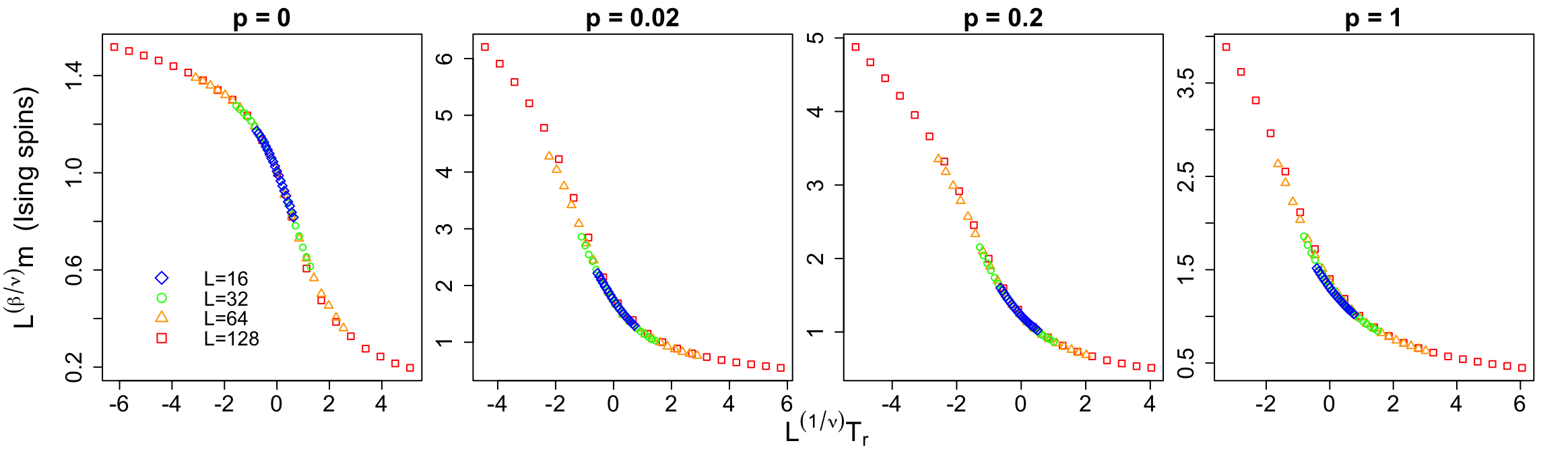}
\caption{\label{fig:finite-size-osc} Finite-size scaling of the order parameter for ecological oscillators (top row) and Ising spins (bottom row). $\beta=1/8$ was used for the $p=0$ plots, and $\beta=1/2$ for the others. The same value of $\nu=1$ was used for all plots following the results from the finite-size scaling of the Binder cumulant (Fig.\ \ref{fig:binder-finite}).}
\end{figure*}

A direct measure of the behavior of the order parameter close to the critical point is given by the exponent $\beta$, which can be obtained by the finite-size scaling of the order parameter, where the magnetization of a system of linear size $L$ is given by 
\begin{equation}
\label{eq:order-finitescaling}
    m_L = L^{-\beta/\nu}F_m(L^{1/\nu}T_r),
\end{equation}
where the function $F_m(x)$ is independent of $L$, such that plots of that function for different system sizes collapse on top of each other if we use appropriate values of $\beta$, $\nu$, and $T_c$ \cite{Newmanbook, landaubinder}. With this method, we find that $\beta=1/8$ and $\nu=1$ suitably fits the regular lattice case for both systems (leftmost plots in Fig.\ \ref{fig:finite-size-osc}), agreeing with 2-D Ising universality class values, and $\beta=1/2$ and $\nu=1$ fits all other values of $p$, agreeing with the mean-field values.

\begin{figure}[ht]
\includegraphics[scale=0.45]{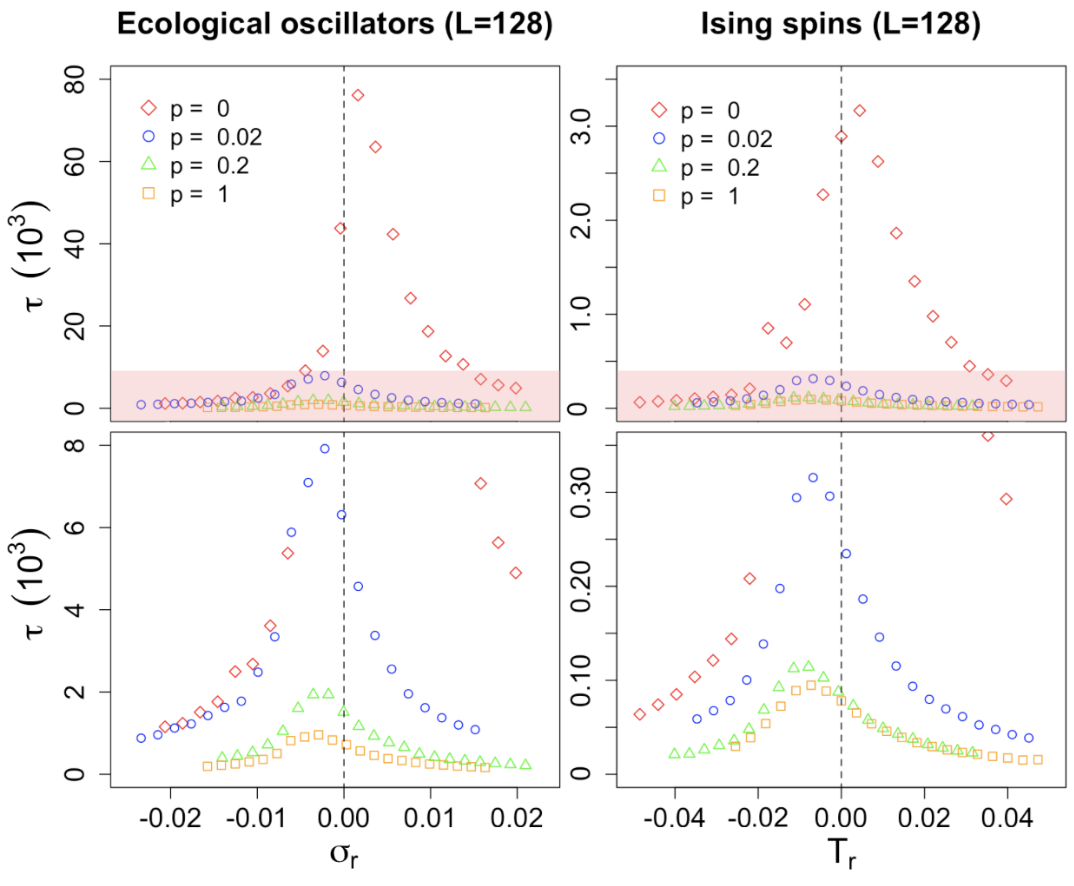}
\caption{\label{fig:autocorrelation-time} Autocorrelation time of the order parameter for ecological oscillators (left) and Ising spins (right). The bottom plots have a rescaled vertical axis {\color{black
}and correspond to the shaded region in the top plots}. The autocorrelation time of ecological oscillators is roughly 20 times greater than the autocorrelation time of Ising spins for the same $p$.}
\end{figure}

{\color{black
}Using the same reduced noise, 
we calculated the integrated autocorrelation time of the order parameter in the stationary state, $\tau=\sum_{t=0}^{T}\rho_m(t)$, where $\rho_m(t)$ is the normalized autocorrelation function of the order parameter \citep{salas1997} and $T$ is an appropriate time lag \citep{grigera2021}} \footnote{{\color{black
}Following \citep{salas1997, grigera2021}, we calculated $\tau$ iteratively. Starting with $T=10$, we set $T=6\tau$ and updated the value of the autocorrelation time until it converged or we summed over the entire length of the autocorrelation function, which indicates a divergence of $\tau$}}. Results are shown in Fig.\ \ref{fig:autocorrelation-time}, where we see that $p=0$ has an autocorrelation time one order of magnitude greater than $p=0.02$ close to criticality for both systems. If $p$ increases to 1, there is a decrease of one more order of magnitude. Furthermore, Ising spins have an autocorrelation time one order of magnitude smaller than ecological oscillators for all values of $p$.

\begin{figure}[ht]
\includegraphics[scale=0.2]{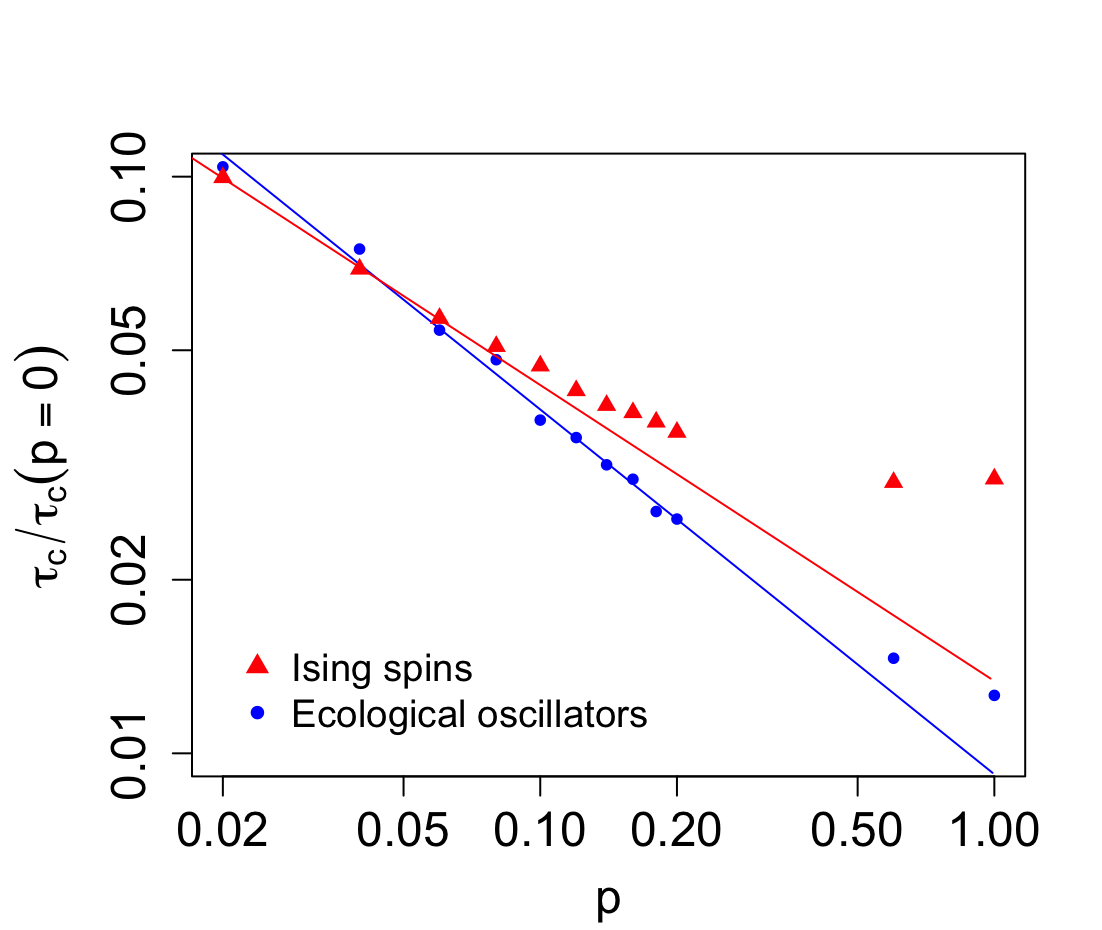}
\caption{\label{fig:autocorrelation-time-powerlaw} Ecological oscillators have power law with exponent $-0.63 \pm 0.02$. Ising spins do not have a clear power law decay; the line shows a power law with exponent $-0.51 \pm 0.02$ which is the best fit for the first three points.}
\end{figure}

Focusing on the critical autocorrelation time, the peaks in Fig.\ \ref{fig:autocorrelation-time}, we observe a power law decay with increasing $p$ for ecological oscillators. For Ising spins, a power law decay is not as clear, but even if present at low values of $p$, the exponent seems to differ from the one for ecological oscillators. These results are presented in Fig.\ \ref{fig:autocorrelation-time-powerlaw}, where the vertical axis shows the critical autocorrelation time normalized by the critical autocorrelation time of the regular lattice.

\begin{figure*}[ht]
\includegraphics[scale=0.55]{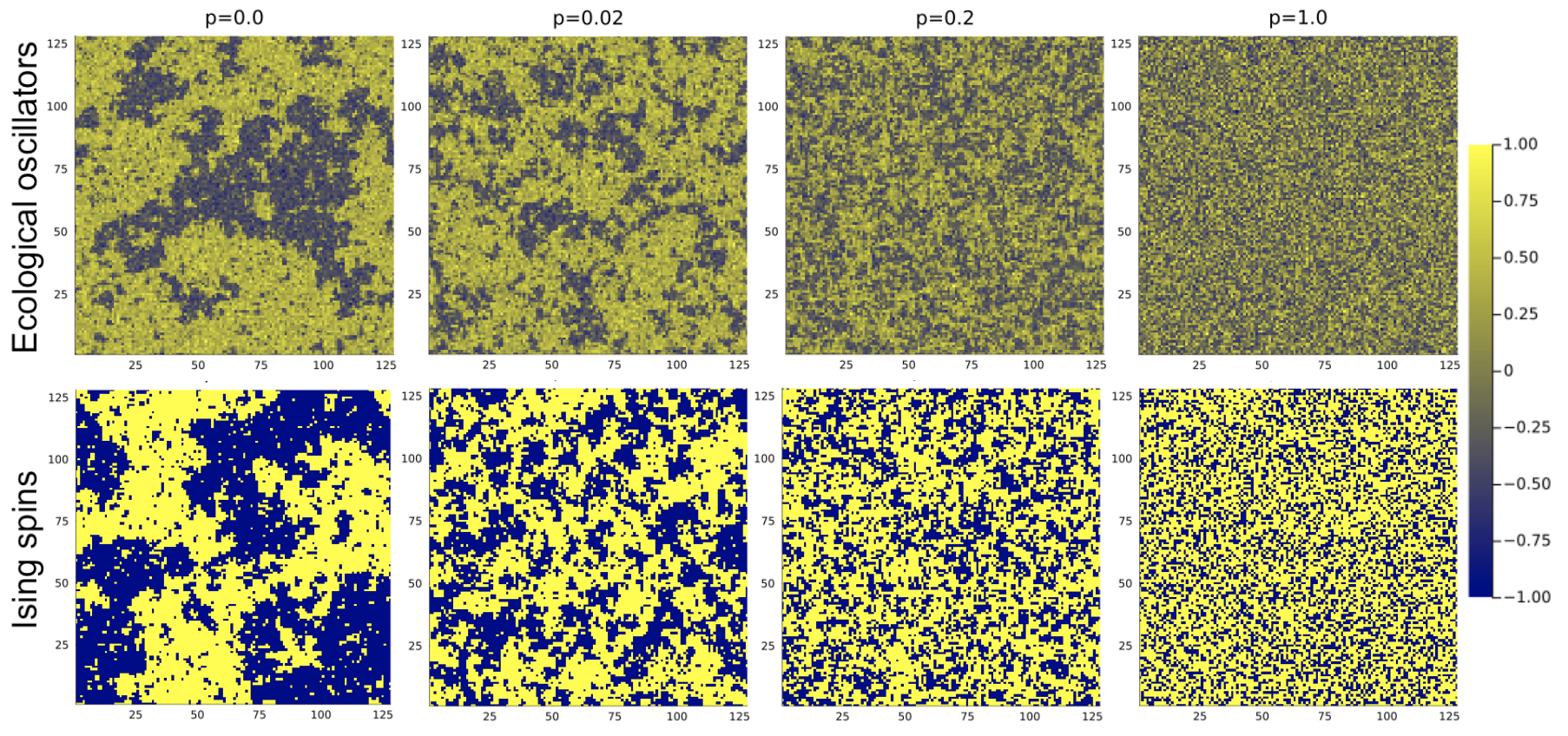}
\caption{\label{fig:snapshots} Snapshots of the two-cycle variable (Eq.\ \eqref{eq:phase-variable}, top row) and Ising spins ($\pm 1$, bottom row) at criticality for different values of $p$. Long-range correlations are broken down as soon as some rewiring is introduced.}
\end{figure*}

\begin{figure}[ht]
\includegraphics[scale=0.65]{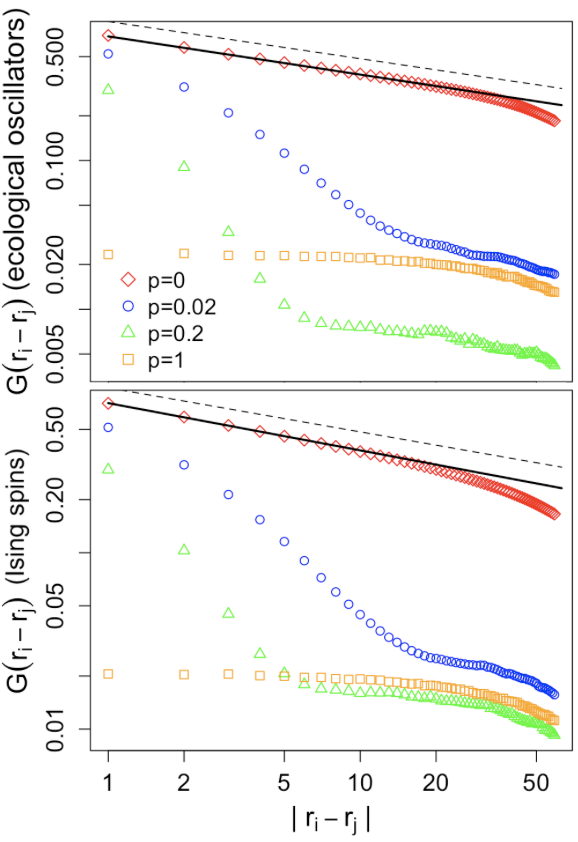}
\caption{\label{fig:correlation-function} {\color{black
}Critical two-point correlation function with $L=128$. The correlation function is measured using only vertical and horizontal spatial lags of up to
$64$ lattice sites.} Results for ecological oscillators are shown on top and for Ising spins at the bottom. The solid black line shows the best power law fit for short distances, with exponent $-0.257 \pm 0.004$ for ecological oscillators and $-0.264 \pm 0.008$ for Ising spins. The dashed line shows a power law with exponent $-0.25$ for reference. The different values of the correlation function at long distances are an effect of the finite size of the system, since $\langle m\rangle$ is not exactly zero at criticality. The decay observed at large distances is an effect of considering only horizontal and vertical correlations.}
\end{figure}

In terms of spatial correlation of the two-cycle variables or Ising spins, introducing long-range interactions quickly breaks down the scale-free pattern observed at criticality. Figure \ref{fig:snapshots} shows snapshots of both systems at one time step at the critical point after the stationary state has been reached. For $p=0$, we observe a fractal pattern with the two-point correlation function decaying as a power law with an exponent equivalent to the expected value of the 2-D Ising universality class, $\eta = 1/4$. For $p>0$, there is no clear power law decay, and the correlation function rapidly decays to a value close to zero, consistent with $\eta=0$ for mean-field. Figure \ref{fig:correlation-function} shows these results, where the two-point correlation function is given by \cite{noble2015, jankebook}

\begin{equation}
\label{eq:correlation-function}
    G(|\mathbf{r_i-r_j}|) = \dfrac{1}{T}\dfrac{1}{N^*}\sum_{t=t_0}^{t_0+T}\sum_{i,j} x_{i,t}x_{j,t},
\end{equation}
where $t_0$ is the burning time until the stationary state is reached, $N^*$ is the number of terms in the inner summation, and $x$ can refer to either the two-cycle variable (Eq.\ \eqref{eq:phase-variable}) or Ising spin. {\color{black
}The vector $\mathbf{r_i}$ is the position vector or site $i$ and the value of $\mathbf{r_i-r_j}$ was varied from $0$ to $L/2$ and taken only in the vertical and horizontal directions.}

\begin{figure}[ht]
\includegraphics[scale=0.2]{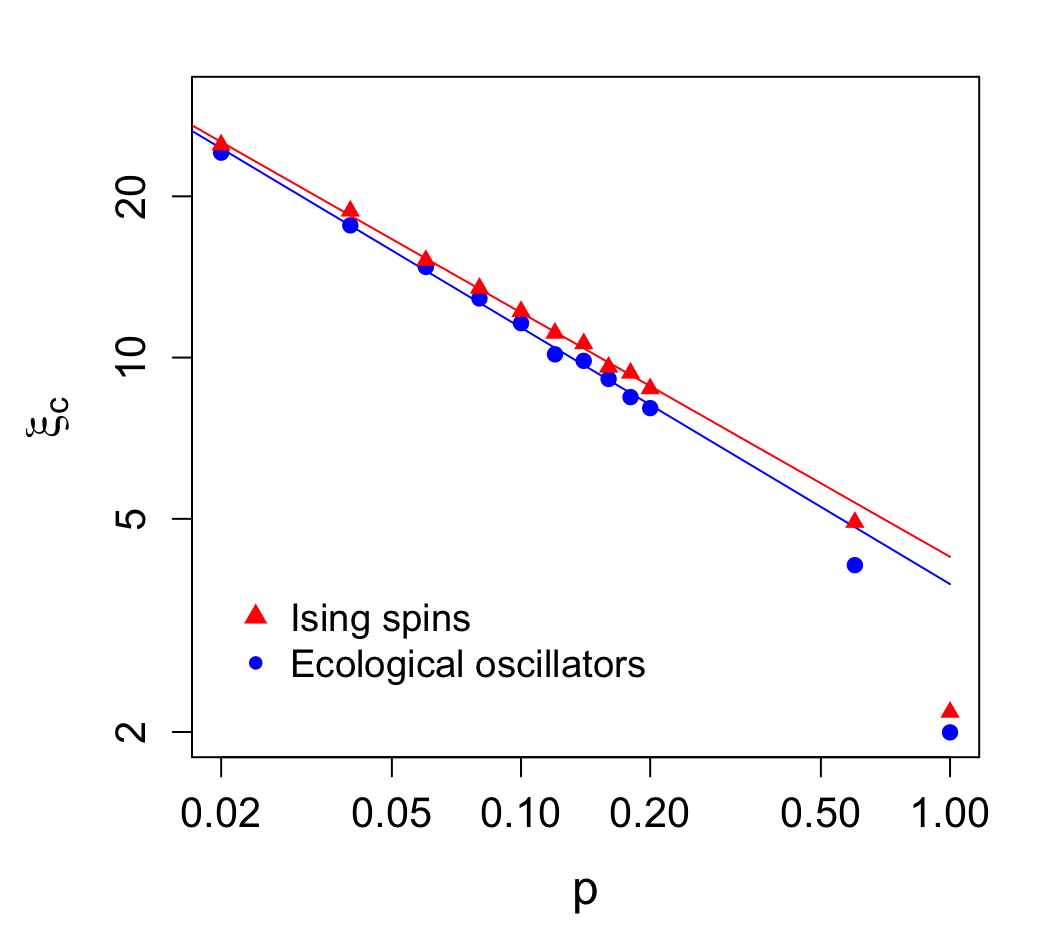}
\caption{\label{fig:correlation-length} The critical correlation length decays with $p$ according to a power law. The lines show the best fit to the first ten points on the graph. The exponent for Ising spins is $-0.46\pm0.01$, and for ecological oscillators it is $-0.48\pm0.01$. The figure shows results for systems with $L=128$. For $p=0$, the correlation length at criticality approaches $L/2$, where the factor of $2$ comes from the periodic boundary conditions.}
\end{figure}

{\color{black
}The correlation function decays exponentially at large distances with a length scale of the correlation length, $\xi$. To determine $\xi$, we first calculate the Fourier transform of the correlation function and fit the points with low wave number to a Lorentzian function, whose coefficients can be combined to obtain $\xi$ \cite{jankebook, grigera2021}}. Fig.\ \ref{fig:correlation-length} shows that the correlation length at criticality decays as a power law for both systems. We note, however, that for $p$ as low as $0.1$, there was no significant dependence of the correlation length with noise, with points being scattered around an average value that decreases with increasing $p$. For $p<0.1$, the correlation length peaks at the critical point, the amplitude of the peak decreases and the curve becomes noisier as we approach $p=0.1$. We also see an important effect of system size which is not captured by finite size scaling. For intermediate values of $p$ ($0.02\leq p\leq 0.2$), we see that, for smaller system sizes ($L=16$ and $32$), the correlation length is a significantly larger fraction of $L$ than for bigger systems ($L=64$ and $128$). This is simply an effect of larger systems having their effective size more drastically reduced by rewiring.

\section{\label{sec:discussion}Discussion}
\subsection{\label{sec:memory} Role of Memory}
As noted in \cite{vahini2020}, ecological oscillators tend to maintain their phase of oscillation, which can be understood in terms of Eq.\ \eqref{eq:ricker}. As a result, ecological oscillators have a longer autocorrelation time than simple Ising spins \cite{vahini2019, vahini2020}. {\color{black
}Although the dynamics of ecological oscillators is Markovian, there is an effective memory of the phase of oscillation that is not present in the standard dynamics of Ising spins.} A quantitative measure of the memory present in the ecological system can be made following the work from Nareddy et al. in \cite{vahini2019, vahini2020}. In \cite{vahini2019}, Nareddy and Machta studied Ising spins with a self-interaction (or memory) term $K$, where a spin $s_j$ flips at time $t$ according not only to its nearest neighbors, but also according to its current state, described by the transition probability \footnote{Note the similarity between this model and the traditional Ising model without memory, which can be recovered if $K=0$.}
\begin{equation}
    p(s_{j,t+1}|t)\propto \exp\left[\left(Ks_{j,t}+J\sum_{\langle k,j\rangle}s_{k,t}\right)s_{j,t+1}\right].
\end{equation}
In the same work, they use the finite-size scaling of the critical autocorrelation time, $\tau = A(K)L^z$ to measure the amplitude of the divergence of the autocorrelation time as a function of memory, finding $A(K)=0.09(1+0.64e^{2K})$ (see Fig.\ 4 in \cite{vahini2019}). From our Fig.\ \ref{fig:autocorrelation-time}, we see that the critical autocorrelation times of ecological oscillators are roughly $20$ times greater than those for Ising spins for most values of $p$. Since our Ising spins have $K=0$, we can use that result to infer that the ecological oscillators have $K\approx 1.9$. This is in almost perfect agreement with the memory value inferred by Nareddy et al.\ for ecological systems (see the rightmost point, which corresponds to the equivalent dispersal rate we use here, in figure 5b in \cite{vahini2020}). For higher values of $p$, the critical autocorrelation time of ecological oscillators decreases more than that of Ising spins. {\color{black
}Naturally, the explanation for this follows the same reasoning as the behavior of the critical temperature, with Ising spins being less sensitive to an increase in the fraction of long range interactions once $p$ reaches a threshold close to $0.2$}. From a memory perspective, this indicates that larger values of $p$ lead to a weaker effect of memory in ecological oscillators. For $p=1$, when the critical autocorrelation time of ecological oscillators is about 10 times that of Ising spins, the value of K would be $K\approx 1.6$. This discrepancy is likely due to the higher values of control parameter for greater values of $p$.

In Fig.\ \ref{fig:phase-diagram}, {\color{black
}we see that both systems display an increase in the critical control parameter as we increase the fraction of long-range coupling, which is the expected facilitation of long-range ordering due to rewiring.} 
A deeper look into those results reveals another display of the memory of ecological oscillators, considering that Ising spins respond more strongly to a small increase in the fraction of long-range interactions. This can be explained by the fact that the dynamics of ecological oscillators depends not only on the nodes to which they are currently coupled, which directly determine the population via Eq.\ \eqref{eq:model}, but also on nodes to which they were previously coupled, which played a role in determining the population at previous time steps and is thus encoded in the oscillator's memory. Averaging a node's connections over time for small $p$ will show us that a node is mostly connected to its nearest neighbors, such that if the system has some memory of its past, it will not respond to rare long-range interactions as dramatically as a system without memory. At high values of $p$, we see another consequence of memory: the critical control parameter remains increasing for ecological oscillators, while it decreases for the Ising model (for $p>0.2$). This effect is perhaps caused by the stochasticity introduced by the rewiring process, since it drives the system out of equilibrium as shown in Fig.\ \ref{fig:balance}, requiring lower temperatures to achieve the ordered state. For ecological oscillators, the critical noise level increases monotonically with $p$, with decreasing slope, because the system has memory, so even with new neighbors at each time step, it is easier to synchronize the network because nodes retain information from their previous connections. This means that the equilibrium state of the ecological system is not as disturbed by rewiring, and is therefore closer to the standard fully connected mean field model. {\color{black
}It would be interesting in future work to carry out a numerical study of the Ising model with memory on rewired networks to confirm that it behaves similarly to ecological oscillators}. A final notable feature shown in Fig.\ \ref{fig:phase-diagram} is that the critical control parameter increases by roughly $25\%$ at its maximum value for both systems. This is significantly less than the increase observed in quenched networks \cite{herrero2002}, and the discrepancy can be understood in terms of the added stochasticity due to the rewiring dynamics.

It is also interesting to notice that the power laws observed in the bottom panels of {\color{black
}Fig.\ \ref{fig:phase-diagram} are compatible with an exponent of $1/2$, which has been previously observed in static small-world networks to be $0.52\pm0.03$ \cite{herrero2002}, although the range of $\log p$ is too small to confirm a true power law. Similar caveats apply to the power laws seen in Figs.\ \ref{fig:autocorrelation-time-powerlaw} and \ref{fig:correlation-length}.}
The similarity emerges despite the different network dynamics, both in terms of local dynamics (ecological oscillators vs Ising spins) and network structure (dynamic vs static rewiring). A satisfactory explanation for such exponent will be the subject of future studies. Similar results have been observed for different systems in different dimensions. Namely, a logarithmic increase in the critical temperature of the XY model was observed when rewiring a 1-D ring lattice \cite{kim2001}, and a power law increase in critical temperature, with exponent $0.96 \pm 0.04$, was observed for the Ising model when rewiring a 3-D lattice \cite{herrero2002}. This allows us to conjecture that the power law may be described by an exponent given by $(d-1)/2$, where $d$ is the dimensionality of the lattice being rewired.

\subsection{\label{sec:universality} Universality Class}
When it comes to universality class arguments, our results show that even a small amount of long-range interaction is sufficient to drive the phase transition into the mean-field universality class. In Fig.\ \ref{fig:critical-binder}, we see that for all $p>0$ the result is consistent with a mean-field transition, and there is no significant difference between the $p=0.02$ and $p=1$ cases. While all points are consistently above the expected value of $U_c=0.27$, this is an effect of the finite size of the network, and Fig.\ \ref{fig:binder-osc} shows that increasing system size leads to slightly lower values of critical Binder cumulant (see the plots for $p=0.02$). {\color{black
} This is due to a crossover between the nearest-neighbor lattice and the mean-field regimes, such that smaller lattices are not significantly impacted by rare long-range interactions.} In the Binder cumulant plots, this makes the apparent value of the critical Binder cumulant closer to the 2-D Ising value for smaller systems. In terms of correlation length, the crossover is also noticeable, since we observe that for $0.02\leq p\leq0.2$ and $L=16$ and $32$, correlation lengths were about $20-40\%$ of $L$, but only $5-15\%$ of $L$ for $L=64$ and $128$. The finite-size scaling analysis of the Binder cumulant results in a critical exponent $\nu=1$ for all values of $p$, and the expected values are $\nu=1$ for the 2-D Ising universality class ($p=0$), and $\nu=1/2$ for the mean-field universality class ($p>0$). While this may seem like a discrepancy, defining a length scale for mean-field and rewired networks is not straightforward \cite{newman1999}. Considering that the number of sites $N$ is the thermodynamic volume of the system, a suitable definition for the linear size of the system is $\ell=N^{1/4}$, where $4$ is the upper critical dimension \cite{ellis2010, kenna2024}. {\color{black
}Using the upper critical dimension is known to yield the correct critical exponents for the complete graph, which is the typical network representation of the mean field case \citep{ellis2010}. For small-world networks, the situation is similar in the sense that average separation between nodes remains finite as $N\rightarrow \infty$.} With that definition, our critical exponent for $p>0$ would be determined, following Eq.\ \eqref{eq:binder-scaling}, by $\Delta U\propto\ell^{1/\nu}=L^{1/(2\nu)}$, such that our results for the appropriate exponent $\nu$ are $1/(2\nu)=1\Rightarrow \nu=1/2$, consistent with the mean-field universality class. The finite-size scaling of the order parameter, shown in
Fig.\ \ref{fig:finite-size-osc}, strengthens the finding that $p>0$ causes the system to be in the mean-field universality class, with a clear collapse of the scaling function when an exponent $\beta=1/2$ is used, as opposed to $\beta=1/8$ for the regular lattice case. The decay of the two-point correlation function in Fig.\ \ref{fig:correlation-function} also supports this finding. Such change in universality class had been observed for similar systems in quenched small-world networks \cite{kim2001, herrero2002, hong2002}.

Finally, an analysis of the correlations in time and space offers some insights on the change of universality class. The significant decrease in autocorrelation time indicates that perturbations to the stationary state dissipate quickly across the entire network, which is a direct effect of the small diameter of small-worlds. Focusing on spatial correlations, the introduction of a small fraction of long-range interactions is enough to break down system-wide correlations, as seen in Figs.\ \ref{fig:snapshots} and \ref{fig:correlation-function}. 

\section{\label{sec:conclusion}Conclusion}
We showed that ecological oscillators and Ising spins have similar qualitative responses to the introduction of occasional random long-range interactions, although, in quantitative terms, Ising spins have a more immediate response to the introduction of such interactions due to their lack of memory, which also leads to a different response to a high fraction of distant coupling. Our results indicate that many quantities display a power law behavior with $p$ at criticality. A theoretical explanation for such behavior as well as a prediction for the exponents remain a subject for further study.

The change in universality class even for small amounts of long-range interactions is an important result for ecology, since it shows that even if long distance migration happens very rarely, it is enough to make the system have the same qualitative behavior as a fully connected ecosystem. Although some systems display spatial patterns that are clearly in the 2-D Ising universality class (for example pistachio orchards \cite{noble2018}), this description seems more useful to metapopulations of sessile individuals that interact via some means other than direct dispersal \cite{shadi2020}. It is reasonable to assume that metapopulations of individuals who actively migrate are expected to experience random long-range migration to some extent, even if small, and such populations will display a phase transition in the mean-field universality class provided that the long-range dispersal happens across a spatial scale comparable to the ecosystem size. From a physics perspective, it is notable that, even though we used a network with dynamic edges, which constantly drives the system out of equilibrium, both systems still manifested the same equilibrium universality as those with static coupling.

To make the model more realistic from a biological perspective, we used a dynamically rewired network, which differs from most models in statistical physics, that use either quenched or annealed networks, both of which satisfy detailed balance. The finding that the dynamic rewiring leads to the same qualitative results as a quenched small-world network is notable, and it is in agreement with previous studies in network theory \cite{lahtinen2002}. A further step to make the model closer to ecological systems would be to introduce directionality of the long-range connections, since there is no reason to assume that long-range migration happens equally in both directions. For physical systems, such directional coupling could be represented by asymmetric or, equivalently, non-reciprocal interactions between spins \cite{aguilera2021}. 

The effect of ecological parameters, such as the growth rate $r$ and the rewiring probability $p$, on the system's memory and autocorrelation time also deserves future attention. The time a population takes to recover from a perturbation is of great significance in ecology, and understanding the results presented in \cite{vahini2020} and here in more detail would represent a significant progress in theoretical ecology.

\begin{acknowledgments}
We thank Benjamin Machta and Gabriel Gellner for fruitful discussions. This work was supported by NSF grant 2325076. 
\end{acknowledgments}

\FloatBarrier

\bibliography{apssamp}

\providecommand{\noopsort}[1]{}\providecommand{\singleletter}[1]{#1}%
\begin{thebibliography}{46}%
\makeatletter
\providecommand \@ifxundefined [1]{%
 \@ifx{#1\undefined}
}%
\providecommand \@ifnum [1]{%
 \ifnum #1\expandafter \@firstoftwo
 \else \expandafter \@secondoftwo
 \fi
}%
\providecommand \@ifx [1]{%
 \ifx #1\expandafter \@firstoftwo
 \else \expandafter \@secondoftwo
 \fi
}%
\providecommand \natexlab [1]{#1}%
\providecommand \enquote  [1]{``#1''}%
\providecommand \bibnamefont  [1]{#1}%
\providecommand \bibfnamefont [1]{#1}%
\providecommand \citenamefont [1]{#1}%
\providecommand \href@noop [0]{\@secondoftwo}%
\providecommand \href [0]{\begingroup \@sanitize@url \@href}%
\providecommand \@href[1]{\@@startlink{#1}\@@href}%
\providecommand \@@href[1]{\endgroup#1\@@endlink}%
\providecommand \@sanitize@url [0]{\catcode `\\12\catcode `\$12\catcode `\&12\catcode `\#12\catcode `\^12\catcode `\_12\catcode `\%12\relax}%
\providecommand \@@startlink[1]{}%
\providecommand \@@endlink[0]{}%
\providecommand \url  [0]{\begingroup\@sanitize@url \@url }%
\providecommand \@url [1]{\endgroup\@href {#1}{\urlprefix }}%
\providecommand \urlprefix  [0]{URL }%
\providecommand \Eprint [0]{\href }%
\providecommand \doibase [0]{https://doi.org/}%
\providecommand \selectlanguage [0]{\@gobble}%
\providecommand \bibinfo  [0]{\@secondoftwo}%
\providecommand \bibfield  [0]{\@secondoftwo}%
\providecommand \translation [1]{[#1]}%
\providecommand \BibitemOpen [0]{}%
\providecommand \bibitemStop [0]{}%
\providecommand \bibitemNoStop [0]{.\EOS\space}%
\providecommand \EOS [0]{\spacefactor3000\relax}%
\providecommand \BibitemShut  [1]{\csname bibitem#1\endcsname}%
\let\auto@bib@innerbib\@empty
\bibitem [{\citenamefont {Strogatz}(2003)}]{strogatz-sync}%
  \BibitemOpen
  \bibfield  {author} {\bibinfo {author} {\bibfnamefont {S.~H.}\ \bibnamefont {Strogatz}},\ }\href@noop {} {\emph {\bibinfo {title} {Sync: The Emerging Science of Spontaneous Order}}}\ (\bibinfo  {publisher} {Hyperion},\ \bibinfo {year} {2003})\BibitemShut {NoStop}%
\bibitem [{\citenamefont {Liebhold}\ \emph {et~al.}(2004)\citenamefont {Liebhold}, \citenamefont {Koenig},\ and\ \citenamefont {Bj{\o}rnstad}}]{liebhold2004}%
  \BibitemOpen
  \bibfield  {author} {\bibinfo {author} {\bibfnamefont {A.}~\bibnamefont {Liebhold}}, \bibinfo {author} {\bibfnamefont {W.~D.}\ \bibnamefont {Koenig}},\ and\ \bibinfo {author} {\bibfnamefont {O.~N.}\ \bibnamefont {Bj{\o}rnstad}},\ }\bibfield  {title} {\bibinfo {title} {Spatial syncrhony in population dynamics},\ }\href@noop {} {\bibfield  {journal} {\bibinfo  {journal} {Annu. Rev. Ecol. Evol. Syst.}\ }\textbf {\bibinfo {volume} {35}},\ \bibinfo {pages} {467} (\bibinfo {year} {2004})}\BibitemShut {NoStop}%
\bibitem [{\citenamefont {Kuramoto}(1975)}]{kuramoto-osc}%
  \BibitemOpen
  \bibfield  {author} {\bibinfo {author} {\bibfnamefont {Y.}~\bibnamefont {Kuramoto}},\ }\bibfield  {title} {\bibinfo {title} {Self-entrainment of a population of coupled non-linear oscillators},\ }in\ \href@noop {} {\emph {\bibinfo {booktitle} {International Symposium on Mathematical Problems in Theoretical Physics}}},\ \bibinfo {editor} {edited by\ \bibinfo {editor} {\bibfnamefont {H.}~\bibnamefont {Arakai}}}\ (\bibinfo  {publisher} {Springer},\ \bibinfo {year} {1975})\ pp.\ \bibinfo {pages} {420--422}\BibitemShut {NoStop}%
\bibitem [{\citenamefont {Moran}(1953)}]{moran1953}%
  \BibitemOpen
  \bibfield  {author} {\bibinfo {author} {\bibfnamefont {P.~A.~P.}\ \bibnamefont {Moran}},\ }\bibfield  {title} {\bibinfo {title} {The statistical analysis of the canadian lynx cycle},\ }\href@noop {} {\bibfield  {journal} {\bibinfo  {journal} {Aust. J. Zool.}\ }\textbf {\bibinfo {volume} {1}},\ \bibinfo {pages} {291} (\bibinfo {year} {1953})}\BibitemShut {NoStop}%
\bibitem [{\citenamefont {Goldwyn}\ and\ \citenamefont {Hastings}(2011)}]{goldwyn2011}%
  \BibitemOpen
  \bibfield  {author} {\bibinfo {author} {\bibfnamefont {E.~E.}\ \bibnamefont {Goldwyn}}\ and\ \bibinfo {author} {\bibfnamefont {A.}~\bibnamefont {Hastings}},\ }\bibfield  {title} {\bibinfo {title} {The roles of the moran effect and dispersal in synchronizing oscillating populations},\ }\href@noop {} {\bibfield  {journal} {\bibinfo  {journal} {J. Theor. Biol.}\ }\textbf {\bibinfo {volume} {289}},\ \bibinfo {pages} {237} (\bibinfo {year} {2011})}\BibitemShut {NoStop}%
\bibitem [{\citenamefont {Noble}\ \emph {et~al.}(2015)\citenamefont {Noble}, \citenamefont {Machta},\ and\ \citenamefont {Hastings}}]{noble2015}%
  \BibitemOpen
  \bibfield  {author} {\bibinfo {author} {\bibfnamefont {A.~E.}\ \bibnamefont {Noble}}, \bibinfo {author} {\bibfnamefont {J.}~\bibnamefont {Machta}},\ and\ \bibinfo {author} {\bibfnamefont {A.}~\bibnamefont {Hastings}},\ }\bibfield  {title} {\bibinfo {title} {Emergent long-range synchronization of oscillating ecological populations without external forcing described by ising universality},\ }\href@noop {} {\bibfield  {journal} {\bibinfo  {journal} {Nat Commun}\ }\textbf {\bibinfo {volume} {6}},\ \bibinfo {pages} {6664} (\bibinfo {year} {2015})}\BibitemShut {NoStop}%
\bibitem [{\citenamefont {Abbott}(2011)}]{abbott2011}%
  \BibitemOpen
  \bibfield  {author} {\bibinfo {author} {\bibfnamefont {K.~C.}\ \bibnamefont {Abbott}},\ }\bibfield  {title} {\bibinfo {title} {A dispersal‐induced paradox: synchrony and stability in stochastic metapopulations},\ }\href@noop {} {\bibfield  {journal} {\bibinfo  {journal} {Ecol. Lett.}\ }\textbf {\bibinfo {volume} {14}},\ \bibinfo {pages} {1158} (\bibinfo {year} {2011})}\BibitemShut {NoStop}%
\bibitem [{\citenamefont {Goldwyn}\ and\ \citenamefont {Hastings}(2008)}]{goldwyn2008}%
  \BibitemOpen
  \bibfield  {author} {\bibinfo {author} {\bibfnamefont {E.~E.}\ \bibnamefont {Goldwyn}}\ and\ \bibinfo {author} {\bibfnamefont {A.}~\bibnamefont {Hastings}},\ }\bibfield  {title} {\bibinfo {title} {When can dispersal synchronize populations?},\ }\href@noop {} {\bibfield  {journal} {\bibinfo  {journal} {Theor. Popul. Biol.}\ }\textbf {\bibinfo {volume} {73}},\ \bibinfo {pages} {395} (\bibinfo {year} {2008})}\BibitemShut {NoStop}%
\bibitem [{\citenamefont {Noble}\ \emph {et~al.}(2018)\citenamefont {Noble}, \citenamefont {Rosenstock}, \citenamefont {Brown}, \citenamefont {Machta},\ and\ \citenamefont {Hastings}}]{noble2018}%
  \BibitemOpen
  \bibfield  {author} {\bibinfo {author} {\bibfnamefont {A.~E.}\ \bibnamefont {Noble}}, \bibinfo {author} {\bibfnamefont {T.~S.}\ \bibnamefont {Rosenstock}}, \bibinfo {author} {\bibfnamefont {P.~H.}\ \bibnamefont {Brown}}, \bibinfo {author} {\bibfnamefont {J.}~\bibnamefont {Machta}},\ and\ \bibinfo {author} {\bibfnamefont {A.}~\bibnamefont {Hastings}},\ }\bibfield  {title} {\bibinfo {title} {Spatial patterns of tree yield explained by endogenous forces through a correspondence between the ising model and ecology},\ }\href@noop {} {\bibfield  {journal} {\bibinfo  {journal} {PNAS}\ }\textbf {\bibinfo {volume} {115}},\ \bibinfo {pages} {1825:1830} (\bibinfo {year} {2018})}\BibitemShut {NoStop}%
\bibitem [{\citenamefont {Hopson}\ and\ \citenamefont {Fox}(2018)}]{fox2018}%
  \BibitemOpen
  \bibfield  {author} {\bibinfo {author} {\bibfnamefont {J.}~\bibnamefont {Hopson}}\ and\ \bibinfo {author} {\bibfnamefont {J.}~\bibnamefont {Fox}},\ }\bibfield  {title} {\bibinfo {title} {Occasional long distance dispersal increases spatial synchrony of population cycles},\ }\href@noop {} {\bibfield  {journal} {\bibinfo  {journal} {J Animal Ecol}\ }\textbf {\bibinfo {volume} {116:903}} (\bibinfo {year} {2018})}\BibitemShut {NoStop}%
\bibitem [{\citenamefont {Kim}\ \emph {et~al.}(2001)\citenamefont {Kim}, \citenamefont {Hong}, \citenamefont {Holme}, \citenamefont {Jeon}, \citenamefont {Minnhagen},\ and\ \citenamefont {Choi}}]{kim2001}%
  \BibitemOpen
  \bibfield  {author} {\bibinfo {author} {\bibfnamefont {B.~J.}\ \bibnamefont {Kim}}, \bibinfo {author} {\bibfnamefont {H.}~\bibnamefont {Hong}}, \bibinfo {author} {\bibfnamefont {P.}~\bibnamefont {Holme}}, \bibinfo {author} {\bibfnamefont {G.~S.}\ \bibnamefont {Jeon}}, \bibinfo {author} {\bibfnamefont {P.}~\bibnamefont {Minnhagen}},\ and\ \bibinfo {author} {\bibfnamefont {M.~Y.}\ \bibnamefont {Choi}},\ }\bibfield  {title} {\bibinfo {title} {{X}{Y} model in small-world networks},\ }\href@noop {} {\bibfield  {journal} {\bibinfo  {journal} {Phys. Rev. E}\ }\textbf {\bibinfo {volume} {64(5)}},\ \bibinfo {pages} {056135} (\bibinfo {year} {2001})}\BibitemShut {NoStop}%
\bibitem [{\citenamefont {Watts}\ and\ \citenamefont {Strogatz}(1998)}]{strogatz1998}%
  \BibitemOpen
  \bibfield  {author} {\bibinfo {author} {\bibfnamefont {D.~J.}\ \bibnamefont {Watts}}\ and\ \bibinfo {author} {\bibfnamefont {S.~H.}\ \bibnamefont {Strogatz}},\ }\bibfield  {title} {\bibinfo {title} {Collective dynamics of ‘small-world’ networks},\ }\href@noop {} {\bibfield  {journal} {\bibinfo  {journal} {Nature}\ }\textbf {\bibinfo {volume} {393}},\ \bibinfo {pages} {440} (\bibinfo {year} {1998})}\BibitemShut {NoStop}%
\bibitem [{\citenamefont {Herrero}(2002)}]{herrero2002}%
  \BibitemOpen
  \bibfield  {author} {\bibinfo {author} {\bibfnamefont {C.~P.}\ \bibnamefont {Herrero}},\ }\bibfield  {title} {\bibinfo {title} {Ising model in small-world networks},\ }\href@noop {} {\bibfield  {journal} {\bibinfo  {journal} {Phys. Rev. E}\ }\textbf {\bibinfo {volume} {66}},\ \bibinfo {pages} {066110} (\bibinfo {year} {2002})}\BibitemShut {NoStop}%
\bibitem [{\citenamefont {Hong}\ \emph {et~al.}(2002)\citenamefont {Hong}, \citenamefont {Choi},\ and\ \citenamefont {Kim}}]{hong2002}%
  \BibitemOpen
  \bibfield  {author} {\bibinfo {author} {\bibfnamefont {H.}~\bibnamefont {Hong}}, \bibinfo {author} {\bibfnamefont {M.~Y.}\ \bibnamefont {Choi}},\ and\ \bibinfo {author} {\bibfnamefont {B.~J.}\ \bibnamefont {Kim}},\ }\bibfield  {title} {\bibinfo {title} {Synchronization on small-world networks},\ }\href@noop {} {\bibfield  {journal} {\bibinfo  {journal} {Phys. Rev. E}\ }\textbf {\bibinfo {volume} {65}},\ \bibinfo {pages} {026139} (\bibinfo {year} {2002})}\BibitemShut {NoStop}%
\bibitem [{\citenamefont {Araujo}\ and\ \citenamefont {de~Aguiar}(2008)}]{araujo2008}%
  \BibitemOpen
  \bibfield  {author} {\bibinfo {author} {\bibfnamefont {S.~B.~L.}\ \bibnamefont {Araujo}}\ and\ \bibinfo {author} {\bibfnamefont {M.~A.~M.}\ \bibnamefont {de~Aguiar}},\ }\bibfield  {title} {\bibinfo {title} {Synchronization and stability in noisy population dynamics},\ }\href@noop {} {\bibfield  {journal} {\bibinfo  {journal} {Phys. Rev. E}\ }\textbf {\bibinfo {volume} {77}},\ \bibinfo {pages} {022903} (\bibinfo {year} {2008})}\BibitemShut {NoStop}%
\bibitem [{\citenamefont {Durrett}\ and\ \citenamefont {Levin}(1994)}]{durrett1994-1}%
  \BibitemOpen
  \bibfield  {author} {\bibinfo {author} {\bibfnamefont {R.}~\bibnamefont {Durrett}}\ and\ \bibinfo {author} {\bibfnamefont {S.~A.}\ \bibnamefont {Levin}},\ }\bibfield  {title} {\bibinfo {title} {Stochastics spatial models: a user's guide to ecological applications},\ }\href@noop {} {\bibfield  {journal} {\bibinfo  {journal} {Phil. Trans. R. Soc. Lond. B}\ }\textbf {\bibinfo {volume} {343}},\ \bibinfo {pages} {329} (\bibinfo {year} {1994})}\BibitemShut {NoStop}%
\bibitem [{\citenamefont {Durrett}\ and\ \citenamefont {Levin}(1999)}]{durrett1999}%
  \BibitemOpen
  \bibfield  {author} {\bibinfo {author} {\bibfnamefont {R.}~\bibnamefont {Durrett}}\ and\ \bibinfo {author} {\bibfnamefont {S.}~\bibnamefont {Levin}},\ }\bibfield  {title} {\bibinfo {title} {Stochastic spatial models},\ }\href@noop {} {\bibfield  {journal} {\bibinfo  {journal} {SIAM}\ }\textbf {\bibinfo {volume} {41}},\ \bibinfo {pages} {677} (\bibinfo {year} {1999})}\BibitemShut {NoStop}%
\bibitem [{\citenamefont {Johst}\ \emph {et~al.}(2002)\citenamefont {Johst}, \citenamefont {Brandl},\ and\ \citenamefont {Eber}}]{johst2002}%
  \BibitemOpen
  \bibfield  {author} {\bibinfo {author} {\bibfnamefont {K.}~\bibnamefont {Johst}}, \bibinfo {author} {\bibfnamefont {R.}~\bibnamefont {Brandl}},\ and\ \bibinfo {author} {\bibfnamefont {S.}~\bibnamefont {Eber}},\ }\bibfield  {title} {\bibinfo {title} {Metapopulation persistence in dynamic landscapes: the role of dispersal distance},\ }\href@noop {} {\bibfield  {journal} {\bibinfo  {journal} {Oikos}\ }\textbf {\bibinfo {volume} {98}},\ \bibinfo {pages} {1263} (\bibinfo {year} {2002})}\BibitemShut {NoStop}%
\bibitem [{\citenamefont {Mart\'{i}}\ and\ \citenamefont {Masoller}(2003)}]{marti2003}%
  \BibitemOpen
  \bibfield  {author} {\bibinfo {author} {\bibfnamefont {A.~C.}\ \bibnamefont {Mart\'{i}}}\ and\ \bibinfo {author} {\bibfnamefont {C.}~\bibnamefont {Masoller}},\ }\bibfield  {title} {\bibinfo {title} {Delay-induced synchronization phenomena in an array of globally coupled logistic maps},\ }\href@noop {} {\bibfield  {journal} {\bibinfo  {journal} {Phys. Rev. E}\ }\textbf {\bibinfo {volume} {67}},\ \bibinfo {pages} {056219} (\bibinfo {year} {2003})}\BibitemShut {NoStop}%
\bibitem [{\citenamefont {de~Roos}\ \emph {et~al.}(1998)\citenamefont {de~Roos}, \citenamefont {McCauley},\ and\ \citenamefont {Wilson}}]{roos1998}%
  \BibitemOpen
  \bibfield  {author} {\bibinfo {author} {\bibfnamefont {A.~M.}\ \bibnamefont {de~Roos}}, \bibinfo {author} {\bibfnamefont {E.}~\bibnamefont {McCauley}},\ and\ \bibinfo {author} {\bibfnamefont {W.~G.}\ \bibnamefont {Wilson}},\ }\bibfield  {title} {\bibinfo {title} {Pattern formation and the spatial scale of interaction between predators and their prey},\ }\href@noop {} {\bibfield  {journal} {\bibinfo  {journal} {Theor. Popul. Biol.}\ }\textbf {\bibinfo {volume} {53}},\ \bibinfo {pages} {108} (\bibinfo {year} {1998})}\BibitemShut {NoStop}%
\bibitem [{\citenamefont {Abbott}\ and\ \citenamefont {Ives}(2012)}]{abbott-enc}%
  \BibitemOpen
  \bibfield  {author} {\bibinfo {author} {\bibfnamefont {K.~C.}\ \bibnamefont {Abbott}}\ and\ \bibinfo {author} {\bibfnamefont {A.~R.}\ \bibnamefont {Ives}},\ }\bibfield  {title} {\bibinfo {title} {Single-species population models},\ }in\ \href@noop {} {\emph {\bibinfo {booktitle} {Encyclopedia of Theoretical Ecology}}},\ \bibinfo {editor} {edited by\ \bibinfo {editor} {\bibfnamefont {A.}~\bibnamefont {Hastings}}\ and\ \bibinfo {editor} {\bibfnamefont {L.~J.}\ \bibnamefont {Gross}}}\ (\bibinfo  {publisher} {University of California Press},\ \bibinfo {year} {2012})\ pp.\ \bibinfo {pages} {1131--1142}\BibitemShut {NoStop}%
\bibitem [{\citenamefont {Ricker}(1954)}]{ricker1954}%
  \BibitemOpen
  \bibfield  {author} {\bibinfo {author} {\bibfnamefont {W.~E.}\ \bibnamefont {Ricker}},\ }\bibfield  {title} {\bibinfo {title} {Stock and recruitment},\ }\href@noop {} {\bibfield  {journal} {\bibinfo  {journal} {J. Fish. Res. Bd. Canada}\ }\textbf {\bibinfo {volume} {11(5)}},\ \bibinfo {pages} {558} (\bibinfo {year} {1954})}\BibitemShut {NoStop}%
\bibitem [{\citenamefont {Bjorkstedt}(2012)}]{ricker-enc}%
  \BibitemOpen
  \bibfield  {author} {\bibinfo {author} {\bibfnamefont {E.~P.}\ \bibnamefont {Bjorkstedt}},\ }\bibfield  {title} {\bibinfo {title} {Ricker model},\ }in\ \href@noop {} {\emph {\bibinfo {booktitle} {Encyclopedia of Theoretical Ecology}}},\ \bibinfo {editor} {edited by\ \bibinfo {editor} {\bibfnamefont {A.}~\bibnamefont {Hastings}}\ and\ \bibinfo {editor} {\bibfnamefont {L.~J.}\ \bibnamefont {Gross}}}\ (\bibinfo  {publisher} {University of California Press},\ \bibinfo {year} {2012})\ pp.\ \bibinfo {pages} {1118--1123}\BibitemShut {NoStop}%
\bibitem [{\citenamefont {Strogatz}(2018)}]{strogatz-nonlinear}%
  \BibitemOpen
  \bibfield  {author} {\bibinfo {author} {\bibfnamefont {S.~H.}\ \bibnamefont {Strogatz}},\ }\href@noop {} {\emph {\bibinfo {title} {Nonlinear Dynamics and Chaos}}}\ (\bibinfo  {publisher} {CRC Press},\ \bibinfo {year} {2018})\BibitemShut {NoStop}%
\bibitem [{\citenamefont {Nareddy}\ \emph {et~al.}(2020)\citenamefont {Nareddy}, \citenamefont {Machta}, \citenamefont {Abbott}, \citenamefont {Esmaeili},\ and\ \citenamefont {Hastings}}]{vahini2020}%
  \BibitemOpen
  \bibfield  {author} {\bibinfo {author} {\bibfnamefont {V.~R.}\ \bibnamefont {Nareddy}}, \bibinfo {author} {\bibfnamefont {J.}~\bibnamefont {Machta}}, \bibinfo {author} {\bibfnamefont {K.~C.}\ \bibnamefont {Abbott}}, \bibinfo {author} {\bibfnamefont {S.}~\bibnamefont {Esmaeili}},\ and\ \bibinfo {author} {\bibfnamefont {A.}~\bibnamefont {Hastings}},\ }\bibfield  {title} {\bibinfo {title} {Dynamical ising model of spatially coupled ecological oscillators},\ }\href@noop {} {\bibfield  {journal} {\bibinfo  {journal} {J. R. Soc. Interface}\ }\textbf {\bibinfo {volume} {17:20200571}} (\bibinfo {year} {2020})}\BibitemShut {NoStop}%
\bibitem [{\citenamefont {Binder}(1981)}]{binder1981}%
  \BibitemOpen
  \bibfield  {author} {\bibinfo {author} {\bibfnamefont {K.}~\bibnamefont {Binder}},\ }\bibfield  {title} {\bibinfo {title} {Finite size scaling analysis of ising model block distribution functions},\ }\href@noop {} {\bibfield  {journal} {\bibinfo  {journal} {Z. Phys. B}\ }\textbf {\bibinfo {volume} {Condensed Matter 43}},\ \bibinfo {pages} {119} (\bibinfo {year} {1981})}\BibitemShut {NoStop}%
\bibitem [{\citenamefont {Landau}\ and\ \citenamefont {Binder}(2009)}]{landaubinder}%
  \BibitemOpen
  \bibfield  {author} {\bibinfo {author} {\bibfnamefont {D.~P.}\ \bibnamefont {Landau}}\ and\ \bibinfo {author} {\bibfnamefont {K.}~\bibnamefont {Binder}},\ }\href@noop {} {\emph {\bibinfo {title} {A Guide to Monte Carlo Simulations in Statistical Physics}}}\ (\bibinfo  {publisher} {Cambridge University Press},\ \bibinfo {year} {2009})\BibitemShut {NoStop}%
\bibitem [{\citenamefont {Newman}\ and\ \citenamefont {Barkema}(2001)}]{Newmanbook}%
  \BibitemOpen
  \bibfield  {author} {\bibinfo {author} {\bibfnamefont {M.~E.~J.}\ \bibnamefont {Newman}}\ and\ \bibinfo {author} {\bibfnamefont {G.~T.}\ \bibnamefont {Barkema}},\ }\href@noop {} {\emph {\bibinfo {title} {Monte Carlo Methods in Statistical Physics}}}\ (\bibinfo  {publisher} {Oxford University Press},\ \bibinfo {year} {2001})\BibitemShut {NoStop}%
\bibitem [{\citenamefont {Taylor}(1997)}]{taylor-book}%
  \BibitemOpen
  \bibfield  {author} {\bibinfo {author} {\bibfnamefont {J.~R.}\ \bibnamefont {Taylor}},\ }\href@noop {} {\emph {\bibinfo {title} {An Introduction to Error Analysis}}}\ (\bibinfo  {publisher} {University Science Books},\ \bibinfo {year} {1997})\BibitemShut {NoStop}%
\bibitem [{\citenamefont {Press}\ \emph {et~al.}(1992)\citenamefont {Press}, \citenamefont {Teukolsky}, \citenamefont {Vetterling},\ and\ \citenamefont {Flannery}}]{num-recipe}%
  \BibitemOpen
  \bibfield  {author} {\bibinfo {author} {\bibfnamefont {W.~H.}\ \bibnamefont {Press}}, \bibinfo {author} {\bibfnamefont {S.~A.}\ \bibnamefont {Teukolsky}}, \bibinfo {author} {\bibfnamefont {W.~T.}\ \bibnamefont {Vetterling}},\ and\ \bibinfo {author} {\bibfnamefont {B.~P.}\ \bibnamefont {Flannery}},\ }\href@noop {} {\emph {\bibinfo {title} {Numerical Recipes in C}}}\ (\bibinfo  {publisher} {Cambridge University Press},\ \bibinfo {year} {1992})\BibitemShut {NoStop}%
\bibitem [{\citenamefont {Chen}\ and\ \citenamefont {Dohm}(2004)}]{chen2004}%
  \BibitemOpen
  \bibfield  {author} {\bibinfo {author} {\bibfnamefont {X.~S.}\ \bibnamefont {Chen}}\ and\ \bibinfo {author} {\bibfnamefont {V.}~\bibnamefont {Dohm}},\ }\bibfield  {title} {\bibinfo {title} {Nonuniversal finite-size scaling in anisotropic systems},\ }\href {https://doi.org/10.1103/PhysRevE.70.056136} {\bibfield  {journal} {\bibinfo  {journal} {Phys. Rev. E}\ }\textbf {\bibinfo {volume} {70}},\ \bibinfo {pages} {056136} (\bibinfo {year} {2004})}\BibitemShut {NoStop}%
\bibitem [{\citenamefont {Janke}\ \emph {et~al.}(1994)\citenamefont {Janke}, \citenamefont {Katoot},\ and\ \citenamefont {Villanova}}]{janke1994}%
  \BibitemOpen
  \bibfield  {author} {\bibinfo {author} {\bibfnamefont {W.}~\bibnamefont {Janke}}, \bibinfo {author} {\bibfnamefont {M.}~\bibnamefont {Katoot}},\ and\ \bibinfo {author} {\bibfnamefont {R.}~\bibnamefont {Villanova}},\ }\bibfield  {title} {\bibinfo {title} {Single-cluster monte carlo study of the ising model on two-dimensional random lattices},\ }\href@noop {} {\bibfield  {journal} {\bibinfo  {journal} {Phys. Rev. B}\ }\textbf {\bibinfo {volume} {49(14)}},\ \bibinfo {pages} {9644} (\bibinfo {year} {1994})}\BibitemShut {NoStop}%
\bibitem [{\citenamefont {Kastening}(2013)}]{kastening2013}%
  \BibitemOpen
  \bibfield  {author} {\bibinfo {author} {\bibfnamefont {B.}~\bibnamefont {Kastening}},\ }\bibfield  {title} {\bibinfo {title} {Anisotropy and universality in finite-size scaling: Critical binder cumulant of a two-dimensional ising model},\ }\href@noop {} {\bibfield  {journal} {\bibinfo  {journal} {Phys. Rev. E}\ }\textbf {\bibinfo {volume} {87}},\ \bibinfo {pages} {044101} (\bibinfo {year} {2013})}\BibitemShut {NoStop}%
\bibitem [{\citenamefont {Parisi}\ and\ \citenamefont {Ruiz-Lorenzo}(1996)}]{parisi1996}%
  \BibitemOpen
  \bibfield  {author} {\bibinfo {author} {\bibfnamefont {G.}~\bibnamefont {Parisi}}\ and\ \bibinfo {author} {\bibfnamefont {J.~J.}\ \bibnamefont {Ruiz-Lorenzo}},\ }\bibfield  {title} {\bibinfo {title} {Scaling above the upper critical dimension in ising models},\ }\href@noop {} {\bibfield  {journal} {\bibinfo  {journal} {Phys. Rev. B}\ }\textbf {\bibinfo {volume} {54(6)}},\ \bibinfo {pages} {R3698} (\bibinfo {year} {1996})}\BibitemShut {NoStop}%
\bibitem [{\citenamefont {Salas}\ and\ \citenamefont {Sokal}(1997)}]{salas1997}%
  \BibitemOpen
  \bibfield  {author} {\bibinfo {author} {\bibfnamefont {J.}~\bibnamefont {Salas}}\ and\ \bibinfo {author} {\bibfnamefont {A.~D.}\ \bibnamefont {Sokal}},\ }\bibfield  {title} {\bibinfo {title} {Dynamic critical behavior of the swendsen-wang algorithm: the two-dimensional three-state potts model revisited},\ }\href@noop {} {\bibfield  {journal} {\bibinfo  {journal} {J. Statist. Phys.}\ }\textbf {\bibinfo {volume} {87}},\ \bibinfo {pages} {1} (\bibinfo {year} {1997})}\BibitemShut {NoStop}%
\bibitem [{\citenamefont {Grigera}(2021)}]{grigera2021}%
  \BibitemOpen
  \bibfield  {author} {\bibinfo {author} {\bibfnamefont {T.~S.}\ \bibnamefont {Grigera}},\ }\bibfield  {title} {\bibinfo {title} {Correlation functions as a tool to study collective behaviour phenomena in biological systems},\ }\href@noop {} {\bibfield  {journal} {\bibinfo  {journal} {J.Phys.Complex.}\ }\textbf {\bibinfo {volume} {2}},\ \bibinfo {pages} {045016} (\bibinfo {year} {2021})}\BibitemShut {NoStop}%
\bibitem [{Note1()}]{Note1}%
  \BibitemOpen
  \bibinfo {note} {{\protect \color {black}Following \protect \citep {salas1997, grigera2021}, we calculated $\tau $ iteratively. Starting with $T=10$, we set $T=6\tau $ and updated the value of the autocorrelation time until it converged or we summed over the entire length of the autocorrelation function, which indicates a divergence of $\tau $}}\BibitemShut {NoStop}%
\bibitem [{\citenamefont {Janke}(2008)}]{jankebook}%
  \BibitemOpen
  \bibfield  {author} {\bibinfo {author} {\bibfnamefont {W.}~\bibnamefont {Janke}},\ }\bibfield  {title} {\bibinfo {title} {Monte carlo methods in classical statistical physics},\ }in\ \href@noop {} {\emph {\bibinfo {booktitle} {Computational Many-Particle Physics}}},\ \bibinfo {editor} {edited by\ \bibinfo {editor} {\bibfnamefont {H.}~\bibnamefont {Fehske}}, \bibinfo {editor} {\bibfnamefont {R.}~\bibnamefont {Schneider}},\ and\ \bibinfo {editor} {\bibfnamefont {A.}~\bibnamefont {Wei{\ss}e}}}\ (\bibinfo  {publisher} {Springer},\ \bibinfo {year} {2008})\ pp.\ \bibinfo {pages} {79--140}\BibitemShut {NoStop}%
\bibitem [{\citenamefont {Nareddy}\ and\ \citenamefont {Machta}(2020)}]{vahini2019}%
  \BibitemOpen
  \bibfield  {author} {\bibinfo {author} {\bibfnamefont {V.~R.}\ \bibnamefont {Nareddy}}\ and\ \bibinfo {author} {\bibfnamefont {J.}~\bibnamefont {Machta}},\ }\bibfield  {title} {\bibinfo {title} {Kinetic ising models with self-interaction: sequential and parallel updating},\ }\href@noop {} {\bibfield  {journal} {\bibinfo  {journal} {Phys. Rev. E}\ }\textbf {\bibinfo {volume} {101}},\ \bibinfo {pages} {012122} (\bibinfo {year} {2020})}\BibitemShut {NoStop}%
\bibitem [{Note2()}]{Note2}%
  \BibitemOpen
  \bibinfo {note} {Note the similarity between this model and the traditional Ising model without memory, which can be recovered if $K=0$.}\BibitemShut {Stop}%
\bibitem [{\citenamefont {Newman}\ and\ \citenamefont {Watts}(1999)}]{newman1999}%
  \BibitemOpen
  \bibfield  {author} {\bibinfo {author} {\bibfnamefont {M.~E.~J.}\ \bibnamefont {Newman}}\ and\ \bibinfo {author} {\bibfnamefont {D.~J.}\ \bibnamefont {Watts}},\ }\bibfield  {title} {\bibinfo {title} {Scaling and percolation in the small-world network model},\ }\href@noop {} {\bibfield  {journal} {\bibinfo  {journal} {Phys. Rev. E}\ }\textbf {\bibinfo {volume} {60}},\ \bibinfo {pages} {7332} (\bibinfo {year} {1999})}\BibitemShut {NoStop}%
\bibitem [{\citenamefont {Ellis}\ \emph {et~al.}(2010)\citenamefont {Ellis}, \citenamefont {Machta},\ and\ \citenamefont {Otto}}]{ellis2010}%
  \BibitemOpen
  \bibfield  {author} {\bibinfo {author} {\bibfnamefont {R.~S.}\ \bibnamefont {Ellis}}, \bibinfo {author} {\bibfnamefont {J.}~\bibnamefont {Machta}},\ and\ \bibinfo {author} {\bibfnamefont {P.~T.-H.}\ \bibnamefont {Otto}},\ }\bibfield  {title} {\bibinfo {title} {Asymptotic behavior of the finite-size magnetization as a function of the speed of approach to criticality},\ }\href@noop {} {\bibfield  {journal} {\bibinfo  {journal} {Ann. Appl. Probab.}\ }\textbf {\bibinfo {volume} {20}},\ \bibinfo {pages} {2118} (\bibinfo {year} {2010})}\BibitemShut {NoStop}%
\bibitem [{\citenamefont {Kenna}\ and\ \citenamefont {Berche}(2024)}]{kenna2024}%
  \BibitemOpen
  \bibfield  {author} {\bibinfo {author} {\bibfnamefont {R.}~\bibnamefont {Kenna}}\ and\ \bibinfo {author} {\bibfnamefont {B.}~\bibnamefont {Berche}},\ }\bibfield  {title} {\bibinfo {title} {Scaling and finite-size scaling above the upper critical dimension},\ }\href@noop {} {\bibfield  {journal} {\bibinfo  {journal} {arXiv:2404.09190}\ } (\bibinfo {year} {2024})}\BibitemShut {NoStop}%
\bibitem [{\citenamefont {Esmaeili}\ \emph {et~al.}(2021)\citenamefont {Esmaeili}, \citenamefont {Hastings}, \citenamefont {Abbott}, \citenamefont {Machta},\ and\ \citenamefont {Nareddy}}]{shadi2020}%
  \BibitemOpen
  \bibfield  {author} {\bibinfo {author} {\bibfnamefont {S.}~\bibnamefont {Esmaeili}}, \bibinfo {author} {\bibfnamefont {A.}~\bibnamefont {Hastings}}, \bibinfo {author} {\bibfnamefont {K.}~\bibnamefont {Abbott}}, \bibinfo {author} {\bibfnamefont {J.}~\bibnamefont {Machta}},\ and\ \bibinfo {author} {\bibfnamefont {V.~R.}\ \bibnamefont {Nareddy}},\ }\bibfield  {title} {\bibinfo {title} {Density dependent resource budget model for alternate bearing},\ }\href@noop {} {\bibfield  {journal} {\bibinfo  {journal} {J. Theor. Biol.}\ }\textbf {\bibinfo {volume} {509}},\ \bibinfo {pages} {110498} (\bibinfo {year} {2021})}\BibitemShut {NoStop}%
\bibitem [{\citenamefont {Lahtinen}\ \emph {et~al.}(2002)\citenamefont {Lahtinen}, \citenamefont {Kert\'{e}sz},\ and\ \citenamefont {Kaski}}]{lahtinen2002}%
  \BibitemOpen
  \bibfield  {author} {\bibinfo {author} {\bibfnamefont {J.}~\bibnamefont {Lahtinen}}, \bibinfo {author} {\bibfnamefont {J.}~\bibnamefont {Kert\'{e}sz}},\ and\ \bibinfo {author} {\bibfnamefont {K.}~\bibnamefont {Kaski}},\ }\bibfield  {title} {\bibinfo {title} {Random spreading phenomena in annealed small world networks},\ }\href@noop {} {\bibfield  {journal} {\bibinfo  {journal} {Physica A}\ }\textbf {\bibinfo {volume} {311}},\ \bibinfo {pages} {571} (\bibinfo {year} {2002})}\BibitemShut {NoStop}%
\bibitem [{\citenamefont {Aguilera}\ \emph {et~al.}(2021)\citenamefont {Aguilera}, \citenamefont {Moosavi},\ and\ \citenamefont {Shimazaki}}]{aguilera2021}%
  \BibitemOpen
  \bibfield  {author} {\bibinfo {author} {\bibfnamefont {M.}~\bibnamefont {Aguilera}}, \bibinfo {author} {\bibfnamefont {S.~A.}\ \bibnamefont {Moosavi}},\ and\ \bibinfo {author} {\bibfnamefont {H.}~\bibnamefont {Shimazaki}},\ }\bibfield  {title} {\bibinfo {title} {A unifying framework for mean-field theories of asymmetric kinetic ising systems},\ }\href@noop {} {\bibfield  {journal} {\bibinfo  {journal} {Nat Commun}\ }\textbf {\bibinfo {volume} {12}},\ \bibinfo {pages} {1197} (\bibinfo {year} {2021})}\BibitemShut {NoStop}%
\end{thebibliography}%

\end{document}